\documentstyle[preprint,eqsecnum,aps]{revtex}

\begin{document}

\draft

\preprint{TPR-96-12; HZPP-96-7; nucl-th/9607044}

\title{Yano-Koonin-Podgoretski\u\i\ Parametrisation \\
       of the Hanbury Brown-Twiss Correlator}

\author{Wu Y.-F.$^{1,}$\cite{address}, U. Heinz$^{1,2}$, 
        B. Tom\'a\v{s}ik$^1$, and U.A. Wiedemann$^1$}

\address{$^1$Institut f\"ur Theoretische Physik, Universit\"at Regensburg,\\
   D-93040 Regensburg, Germany }
\address{$^2$CERN/TH, CH-1211 Geneva 23, Switzerland}

\date{June 11, 1997}

\maketitle

\begin{abstract}

\baselineskip 16pt

The Yano-Koonin-Podgoretski\u\i\ (YKP) parametrisation of Hanbury Brown-Twiss
(HBT) two-particle correlation functions opens new strategies for
extracting the emission duration and testing the longitudinal
expansion in heavy-ion collisions. Based on the recently derived
model-independent expressions, we present a detailed parameter
study of the YKP parameters for a finite, hydrodynamically expanding
source model of heavy-ion collisions. For the class of models studied here,
we show that the three YKP radius parameters have an interpretation as
longitudinal extension, transverse extension and emission duration
of the source in the YKP frame. This frame is specified by
the fourth fit parameter, the Yano-Koonin velocity which describes
to a good approximation the velocity of the fluid element with highest
emissivity and allows to test for the longitudinal expansion of the
source. Deviations from this interpretation of the YKP parameters are
discussed quantitatively. 

\end{abstract}

\pacs{PACS numbers: 25.75.Gz,25.75.Ld,12.38.Mh}

\section{Introduction}
\label{sec1}

The spatio-temporal extension and evolution of the interaction region
in heavy-ion collisions are not directly observable. Indirect experimental
access to its geometry and dynamics is possible through
Hanbury Brown-Twiss (HBT) intensity interferometry~\cite{GKW79,BGJ90}.
However, the interpretation of the measured HBT correlations is in general 
model dependent, and the question arises to what extent their 
interpretational ambiguity can be reduced by a refined analysis of the data.

In general, HBT radius parameters measure the Gaussian widths
(second central moments) of the source distribution in space-time
\cite{CSH95a,CSH95b,CNH95,WSH96,HTWW96}. 
It is the finite lifetime of the particle source in heavy ion 
collisions which complicates their interpretation. For a static 
boson emitting source, the HBT-radii have a {\it unique} interpretation 
in terms of geometrical source sizes. For dynamical sources like those 
created in heavy ion collisions, however, the HBT-radii measure certain 
linear combinations of the lifetime, the geometrical sizes and other
space-time correlations \cite{CSH95a,CSH95b,CNH95,HTWW96,HB95}.
Furthermore, if the source expands 
all HBT parameters become functions of the pair momentum \cite{P84,WSH96}.

For azimuthally symmetric sources, corresponding to heavy-ion
collisions at zero impact parameter, there exist two different
``complete'' Gaussian parametrizations for the correlation function:
the Cartesian parametrization with parameters $R_s$, $R_o$, $R_l$
and $R_{ol}$ \cite{HB95,P84,CSH95a,CSH95b},
and the Yano-Koonin-Podgoretski\u\i\ (YKP) 
parametrisation with parameters $R_{\perp}$, $R_{\parallel}$, $R_0$ 
and $v$~\cite{YK78,P83,GIBS96,CNH95,HTWW96}. In each case,
for expanding sources, these parameters are functions of the pair
momentum. The $R$-parameters have the dimension of a length
while $v$ is a velocity. The Cartesian parametrization has
the additional difficulty that the value and the spatio-temporal
interpretation of its parameters depend strongly on the longitudinal
rest frame of the observer. The YKP radius parameters $R_{\perp}$,
$R_{\parallel}$, and $R_0$, on the other hand, are independent
of the longitudinal velocity of the frame in which the particle
momenta are measured. The fourth YKP fit parameter, the 
Yano-Koonin velocity $v$, singles out a specific longitudinal
rest frame (relative to the observer) in which the spatio-temporal
interpretation of the (longitudinally boost-invariant) YKP-radius
parameters becomes particularly simple. In fact, for ``transparent''
sources (i.e.\ sources, for which particle emission occurs from
the whole volume and is not surface dominated) without
collective transverse expansion, it was shown in \cite{CNH95} that
$R_{\perp}$, $R_{\parallel}$, and $R_0$ give exactly the transverse,
longitudinal and temporal widths respectively, of the source emission
function in the Yano-Koonin frame where $v =0$. We will show
here that in this case, up to small corrections from asymmetries of 
the source rapidity profile due to longitudinal expansion flow, 
$v$ coincides with the longitudinal
velocity of the fluid element around the point of highest emissivity
in the source, such that $R_{\perp}$, $R_{\parallel}$ and $R_0$
measure (and cleanly separate) the transverse, longitudinal and 
temporal lengths of homogeneity of the source in the rest frame
of the emitter. As discussed in Sec.~\ref{sec4a}, the pair momentum
dependence of $v$ allows to measure the longitudinal expansion
of the source in a very direct way.

For the class of models studied here, we will also show (see Sec.~\ref{sec4d})
that in the absence of transverse collective expansion the YKP parameters
show perfect $M_{\perp}$-scaling, i.e., they are, for a given
source, universal functions of the transverse mass $M_{\perp} =
\sqrt{m^2 + {\bf K}_\perp^2}$ of the particle pair, independent
of the particle rest mass. This can be tested by comparing 
$\pi \pi$, $KK$, and $pp$ correlations. The same is not true for
the Cartesian parameters which contain additional kinematic and
frame dependent factors which distinguish between pairs of particles
with different mass. This type of $M_\perp$ scaling of the YKP parameters
is violated for sources with transverse collective expansion; also
other types of transverse $x$-$p$-correlations, like e.g. those
occurring in opaque sources \cite{HV96}, and final state effects
like resonance decays after freeze-out \cite{Bolz}, can break this 
scaling. For a detailed discussion of these effects see Refs.~\cite{TH97}
(opaque sources) and \cite{Heinz96,WH96} (resonance decays).

One of the purposes of this paper is to investigate the specific effects
on the correlator and the related corrections to the spatio-temporal
interpretation of the YKP parameters introduced by transverse expansion 
flow in the source. These will in general depend on the particular
source model, and such an investigation thus must necessarily
involve an extensive model study. We investigate here numerically
a simple parametrization of the source which implements the finite
longitudinal, transverse and temporal extension and the longitudinal
expansion of realistic heavy-ion generated sources and contains
the transverse expansion flow as a tunable parameter. We will see
that strong transverse flow affects the interpretation of the YKP
parameter $R_0$ as the lifetime of the source (in its own rest
frame) and thus renders the extraction of the duration of particle
emission from the correlation data more difficult and less
quantitatively reliable. On the other hand, it also spoils the
$M_{\perp}$-scaling of the YKP radius parameters which opens
the possibility for an independent estimate of the transverse
flow velocity from comparison between pion and kaon correlations.
Clearly, such an estimate will remain somewhat model-dependent,
but the mechanisms isolated in the present paper should still be
very useful for qualitative consistency checks between data and
theoretical interpretation.

The investigation reported here has two aspects: an analytical and a 
numerical one. On the analytical level, we study the connection between
the YKP parameters and the second space-time moments of the emission
function, discussing the dependence of the latter on various geometric
and dynamical features of particle emission. This discussion is largely
model-independent; it is much more detailed than the short account
given in \cite{HTWW96} and should thus serve as a general basis of
understanding which can be used to qualitatively anticipate the
behaviour of the YKP parameters also for other source models than
the one studied here. The numerical side of our study is, of course,
model-dependent and our quantitative results must therefore
be regarded with the necessary caution.

Our paper is organized as follows: In Sec.~\ref{sec2},
we shortly compare the Cartesian and YKP parametrizations, thereby
setting up our notation. In Sec.~\ref{sec3}, we introduce a class of 
hydrodynamical models for the emission function. Sec.~\ref{sec4} contains
a general discussion and a detailed numerical study of these models. 
We focus in particular on the effects of collective expansion flow 
on the YKP parameters. The main results are summarized in  
Sec.~\ref{sec5}.

\section{HBT formalism}
\label{sec2}

We shortly recall the basic relations between the
emission function $S(x,K)$, the measured two-particle correlation 
function $C({\bf q},{\bf K})$, and the different Gaussian parametrisations 
of this correlator in terms of Cartesian or YKP radius parameters. 
We start from the relation \cite{S73,GKW79,P84,CH94} 
(here written down for bosons)
  \begin{equation}
     C({\bf q},{\bf K})
      \approx  1 + {\left\vert \int d^4x\, S(x,K)\,
         e^{iq{\cdot}x}\right\vert^2 \over
         \left\vert \int d^4x\, S(x,K)\right\vert^2 }\, .
  \label{1.1}
  \end{equation}
Here, the emission function $S(x,p)$ is the (Wigner) phase space
density of the boson emitting sources~\cite{S73,P84,CH94} and
denotes the probability that a boson with momentum $p$ is emitted 
from the space time point $x$. It specifies the 
one-particle momentum spectrum $P_1({\bf p}) = E_p dN/d^3p
= \int d^4x\, S(x,p)$ as well as the two-particle correlation 
$C({\bf q},{\bf K})$. The r.h.s. of (\ref{1.1}) has to be evaluated at
$K = \textstyle{1\over 2}(p_1 + p_2)$ (the average momentum of the 
particle pair) and $q = p_1 - p_2$ (their corresponding relative momentum)
where the $p_i$ are on-shell.
The Fourier transform in (\ref{1.1}) does not have a unique
inverse since the four components of the relative momentum $q$
are not independent, due to the on-shell constraint
  \begin{equation}
        q^0 = \bbox{\beta}\cdot {\bf q}\, ,
        \qquad
        \bbox{\beta} = {{\bf K}\over K_0} \approx
        {{\bf K}\over E_K}\, ,
  \label{1.2}
  \end{equation}
which follows from $q{\cdot}K=0$.
In practice the analysis of HBT correlation data must therefore be
based on a comparison with specific models for the emission function
$S(x,K)$, with the aim of constraining the class of ``reasonable"
model sources as far as possible. An important tool for this
procedure are the model-independent expressions for the HBT parameters
\cite{CSH95a,CSH95b,HB95} which allow to calculate from an arbitrary
emission function $S$ the characteristic parameters of the two-particle
correlation function $C$ by simple quadrature.
Experimentally, these HBT parameters are obtained via a
multidimensional Gaussian fit to $C({\bf q},{\bf K})$ in
momentum space. To compute these Gaussian parameters of the
(momentum) correlation function $C$ it is sufficient to use the Gaussian
approximation of the (space-time) emission function $S$,
 \begin{equation}
 \label{1.3}
   S(x,K) = N({\bf K})\, S(\bar x({\bf K}),K)\,
            \exp\left[ - {1\over 2} \tilde x^\mu({\bf K})\,
            B_{\mu\nu}({\bf K})\,\tilde x^\nu({\bf K})\right]
   + \delta S(x,K) \, ,
 \end{equation}
neglecting $\delta S(x,K)$~\cite{WSH96}. Here, the $\tilde{x}_{\mu}$ 
denote the space-time coordinates relative to the effective ``source centre''
$\bar x({\bf K})$ for pions with momentum ${\bf K}$,
 \begin{equation}
 \label{1.4}
  \tilde x^\mu ({\bf K}) = x^\mu - \bar x^\mu({\bf K}) , \qquad
  \bar x^\mu({\bf K}) = \langle x^\mu \rangle , 
 \end{equation}
and 
 \begin{equation}
 \label{1.4a}
  (B^{-1})_{\mu\nu}({\bf K})
  = \langle \tilde x_\mu \tilde x_\nu \rangle
 \end{equation}
is the inverse of the Gaussian curvature tensor in (\ref{1.3}),
adjusted such that the first term in (\ref{1.3}) reproduces the rms width of
the full source $S(x,K)$. The (${\bf K}$-dependent) expectation values in these
definitions are defined as space-time averages over the emission function:
 \begin{equation}
 \label{1.5}
   \langle f(x) \rangle = {\int d^4x \, f(x) \, S(x,K) \over
                              \int d^4x \, S(x,K)} .
 \end{equation}
The correction term $\delta S$ contains information on the
deviation of the emission function $S(x,K)$ from a Gaussian form in coordinate
space, i.e. on sharp edges, wiggles, secondary peaks, etc. 
For the class of models discussed in this paper, however, 
the contributions from $\delta S$ are known to 
have little influence on the half width of the correlation 
function \cite{WSH96}, and can be neglected. Then, the two-particle 
correlation function $C({\bf q},{\bf K})$ can be calculated analytically from
(\ref{1.1}):
 \begin{eqnarray}
 \label{1.6}
  C({\bf q},{\bf K})
    &=& 1 + \exp\left[ - q^\mu q^\nu
            \langle \tilde x_\mu \tilde x_\nu \rangle ({\bf K})
                  \right] \, .
 \end{eqnarray}
It is fully determined by the ${\bf K}$-dependent second space-time
moments $(B^{-1})_{\mu\nu}$ of the source (the ``effective widths"
$\langle \tilde x_\mu \tilde x_\nu \rangle ({\bf K})$ or
``lengths of homogeneity" \cite{CSH95b,AS95}).

\subsection{Gaussian parametrisations of the correlation function}
\label{sec2a}

In general, a Gaussian parametrisation of $C({\bf q},{\bf K})$
is specified by selecting a particular choice of three independent
components of the relative momentum $q$ and implementing in (\ref{1.6})
the on-shell constraint $q{\cdot}K=0$ accordingly. This is usually done in a
Cartesian coordinate system with $z$ along the beam axis and ${\bf K}$
lying in the $x$-$z$-plane. One labels the $z$-component
of a 3-vector by $l$ (for {\em longitudinal}), the $x$-component by
$o$ (for {\em outward}) and the $y$-component by $s$ (for {\em
side-ward}). The mass-shell constraint (\ref{1.2}) reads
 \begin{equation}
 \label{1.7}
   q^0 = \beta_\perp q_o + \beta_l q_l
 \end{equation}
where $\beta_\perp = \vert {\bf K}_\perp \vert / K^0 \approx
\vert {\bf K}_\perp \vert / E_K$ denotes (approximately) the velocity 
of the particle pair transverse to the beam direction, and $\beta_l$ 
its longitudinal component.

The standard Cartesian parametrisation~\cite{CSH95a,CSH95b} of the
correlation function is obtained by using (\ref{1.7}) to eliminate
$q^0$ from Eq.~(\ref{1.6}). This determines 6 Cartesian HBT radius 
parameters $R_{ij}$ in terms of the variances $\langle \tilde x_\mu 
\tilde x_\nu \rangle ({\bf K})$ of the emission function:
 \begin{eqnarray}
    C({\bf q},{\bf K})
    &=& 1 + \exp\left[ -\sum_{i,j=s,o,l} R_{ij}^2({\bf K})\, q_i\, q_j
              \right] \, ,
 \nonumber \\
    R_{ij}^2({\bf K}) &=&
    \langle (\tilde{x}_i-{\beta}_i\tilde{t})
            (\tilde{x}_j-{\beta}_j\tilde{t})\rangle \, ,
    \quad i,j = s,o,l \, .
 \label{1.8}
 \end{eqnarray}
For an azimuthally symmetric collision region, $C({\bf q},{\bf K})$
is symmetric with respect to $q_s \to -q_s$ \cite{CNH95}. Then
$R_{os}^2 = R_{sl}^2 = 0$ and~\cite{CSH95a}
 \widetext
 \begin{equation}
    C({\bf q},{\bf K})
    = 1 + \exp\left[ - R_s^2({\bf K}) q_s^2 - R_o^2({\bf K}) q_o^2
                     - R_l^2({\bf K}) q_l^2 - 2 R_{ol}^2({\bf K}) q_o q_l
              \right] \, ,
 \label{1.9}
 \end{equation}
 \narrowtext
with \cite{CSH95a,HB95}
 \begin{mathletters}
 \label{1.10}
 \begin{eqnarray}
   R_s^2({\bf K}) &=& \langle \tilde{y}^2 \rangle \, ,
 \label{1.10a}\\
   R_o^2({\bf K}) &=&
   \langle (\tilde{x} - \beta_\perp \tilde t)^2 \rangle \, ,
 \label{1.10b}\\
   R_l^2({\bf K}) &=&
   \langle (\tilde{z} - \beta_l \tilde t)^2 \rangle \, ,
 \label{1.10c}\\
   R_{ol}^2({\bf K}) &=&
   \langle (\tilde{x} - \beta_\perp \tilde t)
           (\tilde{z} - \beta_l \tilde t) \rangle \, .
 \label{1.10d}
 \end{eqnarray}
 \end{mathletters}
An alternative way of eliminating the redundant component of
$q$ in (\ref{1.6}) leads to the Yano-Koonin-Podgoretski\u\i\
parametrisation\cite{CNH95,YK78,P83} of $C({\bf q},{\bf K})$,
 \widetext
 \begin{equation}
 \label{1.11}
   C({\bf q},{\bf K}) =
       1 +  \exp\left[ - R_\perp^2({\bf K})\, q_{\perp}^2
                       - R_\parallel^2({\bf K}) \left( q_l^2 - (q^0)^2 \right)
                       - \left( R_0^2({\bf K}) + R_\parallel^2({\bf K})\right)
                         \left(q\cdot U({\bf K})\right)^2
                \right] \, .
 \end{equation}
This is based 
 \narrowtext
on replacing in Eq.~(\ref{1.6}) $q_o$ and $q_s$ in terms
of $q_{\perp} = \sqrt{q_o^2 + q_s^2}$, $q^0$, and $q_l$. Here,
$U({\bf K})$ is a ($K$-dependent) 4-velocity with only a
longitudinal spatial component:
 \begin{equation}
 \label{1.12}
   U({\bf K}) = \gamma({\bf K}) \left(1, 0, 0, v({\bf K}) \right) ,
   \ \ \text{with} \ \
   \gamma = {1\over \sqrt{1 - v^2}}\, .
 \end{equation}
This parametrisation has the advantage that the YKP parameters
$R_\perp^2({\bf K})$, $R_0^2({\bf K})$, and $R_\parallel^2({\bf K})$
extracted from such a fit do not depend on the longitudinal velocity
of the observer system in which the correlation function is measured;
they are invariant under longitudinal boosts. The model-independent
expressions for these YKP-parameters are most conveniently given
in terms of the notational shorthands~\cite{HTWW96}

 \begin{mathletters}
 \label{1.13}
 \begin{eqnarray}
 \label{1.13a}
   A &=& \left\langle \left( \tilde t
         - {\tilde \xi\over \beta_\perp} \right)^2 \right\rangle \, ,
 \\
 \label{1.13b}
   B &=&  \left\langle \left( \tilde z
         - {\beta_l\over \beta_\perp} \tilde \xi \right)^2 \right\rangle
   \, ,
 \\
 \label{1.13c}
   C &=& \left\langle \left( \tilde t - {\tilde \xi\over \beta_\perp} \right)
                      \left( \tilde z - {\beta_l\over \beta_\perp}
                             \tilde \xi \right) \right\rangle \, ,
 \end{eqnarray}
 \end{mathletters}
where $\tilde \xi \equiv \tilde x + i \tilde y$ and $\langle \tilde
y\rangle = \langle \tilde x \tilde y \rangle = 0$ for azimuthally
symmetric sources such that $\langle \tilde \xi^2 \rangle = \langle
\tilde x^2 - \tilde y^2 \rangle$. In terms of these expressions one
finds\footnote{These expressions are valid as long as $(A+B)^2 > 4 C^2$,
i.e. as long as expression (\ref{1.14a}) for the velocity $v$ is defined.
The alternative forms of  $R_\parallel^2$ and $R_0^2$ given in 
\cite{HTWW96} are only valid if additionally $A+B>0$. For a detailed 
discussion see \cite{TH97}.}
 \begin{mathletters}
 \label{1.14}
 \begin{eqnarray}
 \label{1.14a}
   v &=& {A+B\over 2C} \left( 1 - \sqrt{1 - \left({2C\over A+B}\right)^2}
                       \right) \, ,
 \\
 \label{1.14b}
   R_\parallel^2 &=& B - v C,
 \\
 \label{1.14c}
   R_0^2 &=& A - v C ,
 \\
 \label{1.14d}
   R_{\perp}^2 &=& {\langle{ \tilde{y}^2 }\rangle}\, .
 \end{eqnarray}
 \end{mathletters}
For non-vanishing transverse pair momentum $K_{\perp}$, the Cartesian
(\ref{1.8}) and the YKP (\ref{1.11}) parametrisations are
equivalent and it is instructive to compare them. The Cartesian
parameters can be calculated from the YKP ones via \cite{HTWW96}
 \begin{mathletters}
 \label{1.15}
 \begin{eqnarray}
 \label{1.15a}
   R_{\rm diff}^2 &=& R_o^2 - R_s^2 = \beta_\perp^2 \gamma^2
             \left( R_0^2 + v^2 R_\parallel^2 \right) \, ,
 \\
 \label{1.15b}
   R_l^2 &=& \left( 1 - \beta_l^2 \right) R_\parallel^2
             + \gamma^2 \left( \beta_l-v \right)^2
             \left( R_0^2 + R_\parallel^2 \right)\, ,
 \\
 \label{1.15c}
   R_{ol}^2 &=& \beta_\perp \left( -\beta_l R_\parallel^2
             + \gamma^2 \left( \beta_l-v \right)
             \left( R_0^2 + R_\parallel^2 \right) \right)\, ,
 \\
 \label{1.15d}
   R_s^2 &=& R_{\perp}^2\, .
 \end{eqnarray}
 \end{mathletters}
Later we will see that, for the explicit source models studied in this 
paper, in most cases the Yano-Koonin velocity $v$ is very close to the
longitudinal pair velocity $\beta_l$. If this is true (and one should
be careful not to use the following expressions without first checking
this) Eqs.~(\ref{1.15}b,c) simplify to
 \begin{mathletters}
 \label{1.16}
 \begin{eqnarray}
 \label{1.16a}
   R_l^2 &\approx& R_\parallel^2/\gamma^2 \, ,
 \\
 \label{1.16b}
   R_{ol}^2 &\approx& - \beta_\perp \beta_l R_\parallel^2 \, .
 \end{eqnarray}
 \end{mathletters}

There is a slight subtlety for $K_{\perp} = 0$. In this limiting
case, the on-shell constraint (\ref{1.2}) reads
$q^0 = \beta_lq_l$ and cannot be used to eliminate
in (\ref{1.6}) $q_o$ and $q_s$ in terms of $q_{\perp}$,
$q^0$ and $q_3$. Hence, strictly speaking, the YKP parametrisation
exists only for $K_{\perp} \not= 0$. In practice, however, this
does not lead to complications since the $K_{\perp} \to 0$ limit
is well-defined for all YKP-parameters (see Sec.~\ref{sec4b}).

\subsection{Advantages and drawbacks of different Gaussian parametrizations}
\label{sec2b}

The relations (\ref{1.15}) provide a powerful consistency check on 
the experimental fitting procedure of the correlation radii. They 
show that both parametrizations contain exactly the same spatio-temporal
information. However, certain space-time characteristics of the source are 
more directly accessible in a particular parametrization. This is 
especially the case for the duration of the particle emission process, the
``lifetime" of the source.

To see this we return to the expressions (\ref{1.10}) for the Cartesian 
HBT radii. These mix spatial and temporal information on the source in 
a non-trivial and frame-dependent way. Their interpretation in various 
reference systems was analysed analytically \cite{CSH95a,CSH95b,CNH95,WSH96} 
for a large class of (azimuthally symmetric) model emission functions and
compared with the numerically calculated correlation function~\cite{WSH96}.
For these models, the difference
 \begin{equation}
 \label{1.17}
   R_{\rm diff}^2 \equiv  R_o^2 - R_s^2 =
   \beta_\perp^2 \langle \tilde t^2 \rangle - 2 \beta_\perp \langle
   \tilde{x} \tilde t\rangle + \left(\langle \tilde x^2 \rangle -
   \langle \tilde y^2 \rangle\right)
 \end{equation}
is dominated by the first term on the r.h.s. and thus provides access 
to the lifetime $\Delta t = \sqrt{\langle t^2 \rangle - \langle t 
\rangle^2}$ of the source \cite{CP91}. Note, however, that the definitions
(\ref{1.10}) and (\ref{1.17}) are not Lorentz invariant, and that the 
lifetime $\Delta t$ extracted from Eq.~(\ref{1.17}) thus depends on the 
analysis frame. Furthermore, in practice the term $\beta_\perp^2 \langle 
\tilde t^2 \rangle$ turned out to be much smaller than the terms
$\langle \tilde x^2 \rangle$ and $\langle \tilde y^2 \rangle$
which are the leading contributions to $R_o^2$ and $R_s^2$,
respectively \cite{CSH95a,CSH95b,CNH95,WSH96}. As a consequence,
excellent statistics of the data with very small statistical errors 
of $R_o$ and $R_s$ are required to extract the small contribution 
$R_{\rm diff}^2$. This makes the extraction of a small source lifetime 
from the standard fit difficult\footnote{The situation may be better for
very long-lived sources which are predicted by hydrodynamics if there
is a phase transition to a quark-gluon plasma and the collision fireball
is initiated within a certain range of energy densities 
\protect\cite{RG96}.}. Successful attempts have 
been reported from low-energy heavy-ion collisions (using 2-proton 
correlations) where the measured lifetimes are very long: $25\pm 15$ 
fm/$c$ in Ar+Sc collisions at $E/A=80$ MeV \cite{Lisa93} and 
$1400\pm 300$ fm/$c$ in Xe+Al collisions at $E/A=31$ MeV 
\cite{Lisa94} (the latter is the typical evaporation time of a 
compound nucleus). Two-pion correlations at ultra-relativistic
energies ($E/A=200$ GeV) so far failed to yield positive evidence for a
non-vanishing emission duration \cite{NA35,NA44}, except for the heaviest
collision system Pb+Pb \cite{NA49}, but even there the effective lifetime 
is only a few fm/$c$.

For the YKP parametrization the situation is different. The parameters 
$R_0$, $R_\parallel$ and $R_\perp$ are invariant under longitudinal boosts 
and thus independent of the analysis frame. The key to their space-time 
interpretation is provided by the fourth fit parameter $v({\bf K})$. It
specifies a (pair momentum dependent) longitudinal reference frame, 
the Yano-Koonin (YK) frame which defined by $v = 0$ resp. $C=0$ (see
(\ref{1.14a})), in which the space-time variances (\ref{1.14}) for the
YKP radius parameters simplify considerably. Especially, for certain 
classes of source models including the one studied below, the terms 
proportional to $\langle \tilde z \tilde x\rangle$, $\langle \tilde x 
\tilde t \rangle$, and $\langle \tilde x^2 - \tilde y^2 \rangle$ are 
small \cite{CNH95}. Neglecting these terms one obtains \cite{CNH95,HTWW96} 
in the YK frame
 \begin{mathletters}
 \label{1.19}
 \begin{eqnarray}
 \FL
   R_\perp^2({\bf K}) &=& \langle \tilde{y}^2 \rangle \, ,
 \label{1.19a} \\
   R_\parallel^2({\bf K}) &=& B =
   \left\langle \left( \tilde z - {\beta_l\over\beta_\perp} \tilde x
                \right)^2 \right \rangle
     - {\beta_l^2\over\beta_\perp^2} \langle \tilde y^2 \rangle
    \approx \langle \tilde z^2 \rangle \, ,
 \label{1.19b} \\
   R_0^2({\bf K}) &=& A =
   \left\langle \left( \tilde t - {1\over\beta_\perp} \tilde x
                \right)^2 \right \rangle
   - {1\over\beta_\perp^2} \langle \tilde y^2 \rangle
   \approx \langle \tilde t^2 \rangle \, .
 \label{1.19c}
 \end{eqnarray}
 \end{mathletters}
In neighbouring frames, the Yano-Koonin velocity can be calculated 
with the same approximations as
 \begin{equation}
 \label{1.18}
   v \approx {C\over A+B}
     \approx {\langle \tilde z \tilde t \rangle \over
              \langle \tilde t^2 \rangle + \langle \tilde z^2 \rangle}
 \, .
 \end{equation}
Note that in the YK frame the temporal structure of the source 
enters only in the parameter $R_0$. Its leading conribution is 
given by the time $\Delta t({\bf K})=\sqrt{\langle \tilde t^2 \rangle}$ 
during which particles of momentum ${\bf K}$ are emitted in this frame.
In the YKP parametrisation $R_0 \approx \Delta t$ is fitted directly and 
not obtained as the difference of two large fit parameters as in the 
Cartesian fit.

In practice, however, the extraction of $R_0$ from the YKP fit 
is still not easy. From Eq.~(\ref{1.11}) it follows that in the YK frame
$R_0$ must be extracted from the $q^0$-dependence of the correlator. 
Due to the mass-shell constraint (\ref{1.2}) the interesting range of $q^0$ 
is limited, especially for low-momentum pairs, and the sensitivity of
the fit function (\ref{1.11}) to $R_0$ is weaker than to the two other 
radius parameters. $R_0$ values thus tend to come out with larger 
experimental error bars.

\section{A model for a finite expanding source}
\label{sec3}

For our numerical study we have taken the model from Ref.~\cite{CNH95}
with the emission function
 \begin{equation}
   S(x,K) = {M_\perp \cosh(\eta-Y) \over
            (2\pi)^3 \sqrt{2\pi(\Delta \tau)^2}}
        \; \exp \left[- {K \cdot u(x) \over T}\right]
        \; \exp \left[- {(\tau-\tau_0)^2 \over 2(\Delta \tau)^2}
                      - {r^2 \over 2 R^2}
                      - {(\eta- \eta_0)^2 \over 2 (\Delta \eta)^2}
                \right] .
 \label{2.1}
 \end{equation}
The first term specifies the shape of the freeze-out hypersurface, 
the second one is a Lorentz-covariant Boltzmann factor
encoding the assumption of local thermal equilibration
superimposed by collective expansion,
while the last one has a purely geometrical interpretation.
The space-time coordinates in longitudinal and temporal directions 
are parametrised by the space-time rapidity $\eta = {1 \over 2}
\ln[(t+z)/(t-z)]$ and the longitudinal proper time $\tau= \sqrt{t^2-
z^2}$. In the transverse direction, the radius is $r = \sqrt{x^2+y^2}$.
Accordingly, the measure reads
$d^4x = \tau\, d\tau\, d\eta\, r\, dr\, d\phi$.
The time--component of the pair momentum is set to the on-shell
value $K_0 = E_K = \sqrt{m^2 + {\bf K}^2}$. This approximation
was studied in detail in Ref.\cite{CSH95b} where it was shown to
be acceptable. Thus the pair momentum $K$ can be parametrised
using the momentum rapidity $Y = {1\over 2}
\ln[(1+\beta_l)/(1-\beta_l)]$ and the transverse mass
$M_\perp = \sqrt{m^2 + K_\perp^2}$,
  \begin{equation}
    \label{2.2}
    K^\mu = (M_\perp\cosh{Y}, K_\perp, 0, M_\perp\sinh{Y}).
  \end{equation}
We implement longitudinal and azimuthally symmetric transverse expansion 
of the source by parametrising the flow velocity in the form
\begin{equation}
 \label{2.3}
   u^{\mu}(x) = \left( \cosh \eta_l(\tau,\eta) \cosh \eta_t(r), \,
                     {x\over r}\sinh \eta_t(r),  \,
                     {y\over r}\sinh \eta_t(r),  \,
                     \sinh \eta_l(\tau,\eta) \cosh \eta_t(r) \right) .
 \end{equation}
For the longitudinal flow rapidity we take $\eta_l(\tau,\eta) = \eta$ 
independent of $\tau$, i.e. we assume a Bjorken scaling profile \cite{BJ83}
$v_l=z/t$ in the longitudinal direction. For the transverse flow rapidity 
we take a linear profile of strength $\eta_f$:
 \begin{equation}
 \label{2.4}
  \eta_t(r) = \eta_f \left( {r \over R} \right)\, .
 \end{equation}
The scalar product in the exponent of the Boltzmann factor generates
the $x$-$K$-correlation in our source. It can then be written as
  \begin{equation}
    \label{2.5}
    K\cdot u(x) = M_\perp \cosh(\eta - Y) \cosh\eta_t(r) 
                - K_\perp {x\over r} \sinh\eta_t(r) \, ,
  \end{equation}
Please note that for non-zero transverse momentum $K_\perp$, a finite 
transverse flow breaks the azimuthal symmetry of the emission function 
via the second term in (\ref{2.5}). For $\eta_f=0$, the emission function
is azimuthally symmetric for all $K_{\perp}$. Also, it then has no 
explicit $K_\perp$-dependence, and $M_\perp$ is the only relevant scale. 
As will be discussed in Sec.~\ref{sec4d} this gives rise to perfect 
$M_\perp$-scaling of the YKP radius parameters in the absence of 
transverse flow, which is again broken for non-zero transverse flow 
\cite{HTWW96a}. 

Besides $\eta_f$, the model parameters are the freeze-out temperature 
$T$, the transverse geometric (Gaussian) radius $R$, the average 
freeze-out proper time $\tau_0$ as well as the mean proper emission 
duration $\Delta \tau$, the centre of the source rapidity distribution 
$\eta_0$, and the (Gaussian) width of the space-time rapidity profile 
$\Delta \eta$. A rough spatial picture of the source at various fixed 
coordinate times can be gleaned from the Figs.\ 1 and 2 in Ref.~\cite{CN96} 
(although their source has sharp edges whereas ours is smoothed by 
Gaussian profiles) and from Figs.~1 and 2 in Ref.~\cite{TH97}. 
Note that our parametrization of $S(x,K)$ does not allow for the case 
of opaque sources where the emission is surface dominated~\cite{HV96,TH97}. 
In this case, the contribution of ${\langle{ \tilde{x}^2 - \tilde{y}^2 
}\rangle}$ may become negative and large \cite{HV96} which might alter 
the argument following (\ref{1.17}). A detailed study of such opaque
sources is presented in Ref.~\cite{TH97}. On the basis of a comparison 
with the preliminary data of Ref.~\cite{NA49} the authors of that study 
conclude that the source created in Pb+Pb collisions at the CERN SPS
is rather transparent and not opaque; its qualitative features are well
described by the model presented here \cite{H97}.

We did our calculations for pions ($m=m_{\pi^\pm}=139$ MeV/$c^2$)
and kaons ($m=m_{K^\pm}=494$ MeV/$c^2$). Resonance decays 
\cite{Bolz,Heinz96} are not discussed here but is deferred 
to a separate publication \cite{WH96}. The calculations presented here
are meant to illustrate general properties of the YKP parameters; no
attempts to compare with data will be made.

\section{Lifetimes and sizes from the YKP-fit to the
         correlation function}
\label{sec4}

In this section we present a quantitative study of the YKP fit-parameters. 
Since the YKP-parameter 
$R_{\perp}$ is identical to the ``side'' radius of the Cartesian
parametrisation, $R^2_{\perp} = R^2_s = {\langle{y^2}\rangle}$,
its interpretation is obvious and independent of the longitudinal 
velocity of the reference frame. Hence we focus subsequently on the 
remaining three parameters $R^2_0$, $R^2_\parallel$, and $v$.

Unless stated otherwise, the numerical calculations below
are done with the set of source parameters $T = 140$ MeV, $R = 3$ fm,
$\Delta \eta = 1.2$, $\tau_0 = 3$ fm/$c$, $\Delta\tau = 1$ fm/$c$.

\subsection{The Yano-Koonin velocity}
\label{sec4a}

According to Eq~(\ref{1.12}), the YKP fit parameter $v$ is a
longitudinal velocity. In this subsection we give a detailed
discussion of the reference frame 
specified by $v$ and establish its relation to several other 
commonly used reference frames. Their definitions are:
\begin{itemize}
\item
  {\bf CMS}: The centre of mass frame of the fireball, specified by
        $\eta_0 = 0$.
\item
  {\bf LCMS} (Longitudinally CoMoving System \cite{CP91}): A pion 
        (kaon) pair-dependent frame, specified by $\beta_l = Y =0$. 
        In this frame, only the transverse velocity component of the 
        pion (kaon) pair is non-vanishing.
\item
  {\bf LSPS} (Longitudinal Saddle-Point System \cite{CL96}):
        The longitudinally moving rest frame of the point of
        maximal emissivity for a given pair momentum. In general, 
        the velocity of this frame depends on the momentum 
        of the emitted particle pair. For symmetric sources 
        the point of maximal emissivity (``saddle point'')
        coincides with the ``source centre'' $\bar x({\bf K})$
        defined in (\ref{1.4}). In this approximation, for a 
        source like (\ref{2.1}), the LSPS velocity is given
        by the longitudinal component of $u^\mu(\bar x({\bf K}))$.
\item
  {\bf YK} (Yano-Koonin frame \cite{HTWW96}): The frame for which the 
        YKP velocity parameter vanishes, $v({\bf K})=0$. Again, this 
        frame is in general pair momentum dependent.
\end{itemize}

These four frames are quite different in nature. The velocities (or 
rapidities) of the CMS and LCMS frames can be easily determined 
experimentally, the first from the peak in the single particle rapidity 
distribution, the second from the longitudinal momentum of the measured
pion pair. However, the velocity of the LSPS is in a sense more interesting: 
from its definition it is directly related to the longitudinal expansion 
velocity of the source. 
In fact, longitudinal expansion of the source leads to a characteristic
dependence of the LSPS velocity $v_{_{\rm LSPS}}$ on the pair rapidity.
This is most easily seen by considering two extreme fireball models:
 
(1) If the source does not expand, all its elements move with the same 
velocity (rapidity), namely that of the CMS. Then there is no 
kinematic difference between different parts of the fireball, and
the saddle point is $K$-independent and given by the peak of the
space-time distribution of the source. Thus the rapidity of the LSPS, 
defined as $Y_{_{\rm LSPS}} = \frac 12 \ln \left[(1+v_{_{\rm LSPS}})
/(1-v_{_{\rm LSPS}})\right]$, is {\it independent} of the pair rapidity 
$Y$ and identical to the rapidity of the CMS.
 
This behaviour of $Y_{_{\rm LSPS}}$ arises automatically if
the source has no $x - K$ correlations, i.e. if the emission function 
factorizes, $S(x,K)=F(x)\, G(K)$. Then all space-time characteristics 
of the source, including the LSPS velocity, are determined by $F(x)$ 
alone and do not depend on $K$. Hence $Y_{_{\rm LSPS}}$ is independent of 
$Y$. Factorization is, however, not necessary for this behaviour: 
Non-vanishing $x-K$ correlations generated, e.g., by temperature 
gradients don't induce a $Y$-dependence of $Y_{_{\rm LSPS}}$ as
long as the source does not expand longitudinally.

(2) If we set in Eq.~(\ref{2.1}) $\Delta\eta\to\infty$, we recover
the Bjorken model \cite{BJ83} for a longitudinally infinite source 
with boost-invariant longitudinal expansion. The only $\eta$-dependence
then comes from the thermal Boltzmann factor, and the longitudinal saddle 
point obviously lies at $\eta=Y$. Since for the Bjorken scaling profile
$\eta$ coincides with the longitudinal fluid rapidity, this implies
that in this case the rapidity of the LSPS is {\it identical} with the 
pair rapidity, $Y_{_{\rm LSPS}}=Y$, i.e. the LSPS coincides with the LCMS. 

It is obvious from this discussion that knowledge of the function 
$Y_{_{\rm LSPS}}(Y)$ would allow to distinguish between these two 
scenarios: A non-expanding source would yield $Y_{_{\rm LSPS}}(Y)=
{\rm const.}$ while a source with boost-invariant longitudinal expansion 
gives $Y_{_{\rm LSPS}}(Y)=Y$. Realistic models are expected to lie in 
between these two extremes. 

Unfortunately, the LSPS-velocity $v_{_{\rm LSPS}}({\bf K})$ cannot be 
measured. This is clear from its definition as the longitudinal flow 
velocity evaluated at the point of maximum emissivity. As discussed 
above, this point is approximately given by the source centre 
$\bar x({\bf K})$ which itself is unmeasurable: it drops out 
\cite{HTWW96} from both the single and two-particle spectra which 
are invariant under a translation of the source centre (even if 
${\bf K}$-dependent!). The only velocity we can measure, namely from 
an YKP fit of the two-particle correlator, is the Yano-Koonin (YK) 
velocity $v({\bf K})$ resp. the associated rapidity $Y_{_{\rm YK}} 
= \frac 12 \ln \left[(1+v)/(1-v)\right]$. It is therefore 
very gratifying to know that for sufficiently rapidly expanding systems
the two velocities are always closely related (although in general not 
identical). In particular, we will show below that for the class of models 
(\ref{2.1}) the function $Y_{_{\rm YK}}(Y)$ shares with $Y_{_{\rm LSPS}}(Y)$ 
the feature that it provides a direct signature for longitudinal expansion. 

The close relationship between the (measurable) YK-velocity and the
(theoretical) LSPS-velocity for certain source models has been known 
for years. In Ref.~\cite{P83} a ``symmetric frame'' was introduced as 
the reference frame, in which the production process is symmetric in the beam
direction. In this frame, the longitudinal extension and the lifetime of 
the source reach their extremal values~\cite{P83}. For moving, but 
non-expanding azimuthally symmetric sources Podgoretski\u\i\ found
in this way the parametrisation (\ref{1.11}), with ${\bf K}$-independent
parameters $R_\perp$, $R_\parallel$, and $R_0$, and with what we call
the Yano-Koonin velocity $v$ being identical to the velocity of his 
``symmetric frame''. In this case the YK-system also coincides with 
the rest frame of the source as a whole (CMS) as well as with the LSPS.

That the coincidence between the YK and LSPS systems is more generally 
valid for sources which are
symmetric around their saddle point $\bar x({\bf K})$ has been observed 
in Refs.~\cite{CNH95,CL96}. It thus holds for any emission function 
in the Gaussian saddle-point approximation, due to the symmetry of the 
latter. As the following paragraph will show, differences between the 
YK-velocity and the velocity of the LSPS are only due to asymmetries 
of the source around its saddle point. Although such 
asymmetries usually exist for collectively expanding sources with finite
geometric extension, they are generally small and can be treated
perturbatively. Therefore, the YK-frame and the LSPS-frame are usually 
very close to each other, $v \approx v_{_{\rm LSPS}}$. From the examples 
above it is clear, however, that the same is not true for the LCMS (i.e. the 
longitudinal rest frame defined by the pair rapidity $Y$), and that 
generally $v \ne v_{_{\rm LCMS}}$.

Let us now discuss the difference $v - v_{_{\rm LSPS}}$ in more detail.
If it is small, so is $C$ when evaluated in the LSPS-frame. From 
Eqs.~(\ref{1.18}) and (\ref{1.13c}) we see that then in the LSPS frame 
$v$ is given by
 \begin{equation}
   v \approx  \frac C{A+B}
     \approx  {1 \over  \langle \tilde z^2 \rangle
                      + \langle \tilde t^2 \rangle}
     \left( \langle \tilde z \tilde t \rangle 
          - \frac {1}{\beta_\perp} \langle \tilde z \tilde x \rangle
          - \frac {\beta_l}{\beta_\perp} \langle \tilde t \tilde x \rangle
          + \frac {\beta_l}{\beta_\perp^2}
            \langle \tilde x^2 - \tilde y^2 \rangle 
            \right) \, ,
 \label{3.1}
\end{equation}
where we expanded in first order of $\langle 
\tilde t \tilde x \rangle$, $\langle \tilde z \tilde x \rangle$, 
$\langle \tilde x^2 - \tilde y^2 \rangle$. The smallness of these 
terms was argued in \cite{CNH95} and will be checked in the following 
subsection. The first and second term of (\ref{3.1}) reflect the 
longitudinal asymmetry of the source around the saddle point; they 
vanish for sources with longitudinal reflection symmetry $\tilde z 
\to -\tilde z$. Similarly, the second and third term vanish unless 
the reflection symmetry $\tilde x \to -\tilde x$ around the saddle 
point in the ``out"-direction is broken (e.g. by transverse flow). 
(Asymmetry in $t$ is needed for the first and third terms to become 
non-zero.) A non-zero value of the last term, finally, 
indicates the breaking of the ``out"-``side" rotation symmetry; this
can again be caused by transverse source gradients as e.g. transverse 
flow (see Eq.~(\ref{2.5})). We conclude that the difference 
$v - v_{_{\rm LSPS}}$ is entirely due to
asymmetries of the source around the saddle point. Furthermore, we
will show in Sec.~\ref{sec4b} that the last three terms in (\ref{3.1}) 
are small for small values of the transverse flow rapidity $\eta_f$ and/or
small values of $K_\perp$. (They vanish for $\eta_f = 0$.)
In these limits the difference between $v$ 
and $v_{_{\rm LSPS}}$ is dominated by the longitudinal source asymmetry,
and $v$ is very accurately given by the (leading) first term in (\ref{3.1}),
see Eq.~(\ref{1.18}). Note that in our model the breaking of the
longitudinal reflection symmetry is due to the non-symmetric rapidity 
profile of the emission function for $Y \ne \eta_0$.

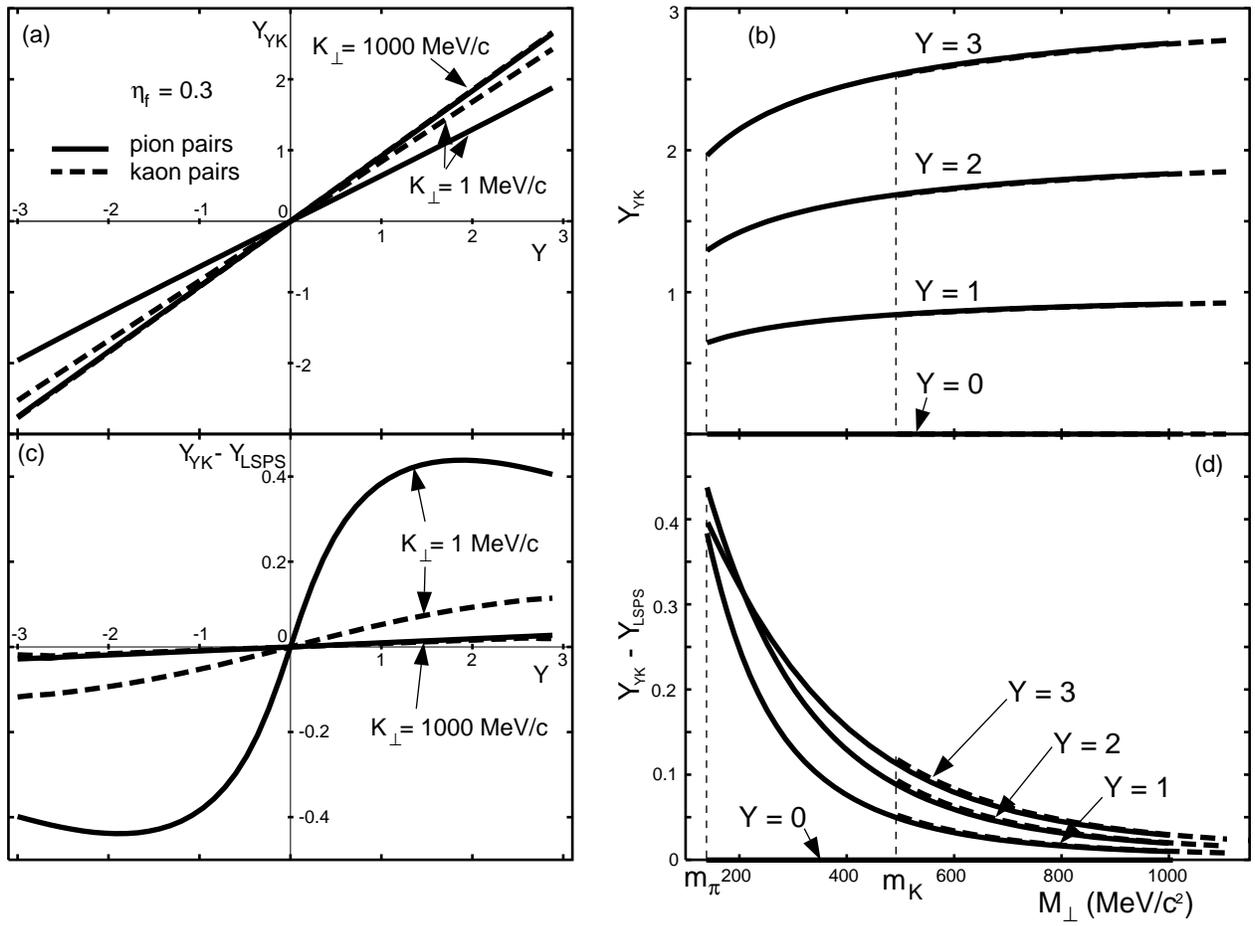
\begin{figure}
\caption{
 (a) The rapidity of the YK frame as a function of the
     pair rapidity $Y$ (both measured in the source CMS), for
     pions (solid) and kaons (dashed) and for transverse momenta 
     $K_\perp=1$ MeV and $K_\perp=1000$ MeV. The transverse flow 
     was set to $\eta_f = 0.3$.
 (b) Same as (a), but shown as a function of $M_\perp$ for different
     values of $Y$. 
 (c) The difference $Y_{_{\rm YK}} - Y_{_{\rm LSPS}}$ (see text),
     plotted in the same way as (a).
 (d) Same as (c), but shown as a function of $M_\perp$ for different
     values of $Y$.
}
\label{F1}
\end{figure}

For a quantitative discussion we plot in Fig.~\ref{F1} the Yano-Koonin 
rapidity $Y_{_{\rm YK}}$
as well as the difference $Y_{_{\rm YK}} - Y_{_{\rm LSPS}}$ as functions of 
$M_\perp$ and $Y$. All rapidities are given relative to the CMS. One sees
that for large values of $K_\perp$ the agreement of $Y_{_{\rm YK}}$ with
$Y_{_{\rm LSPS}}$ is almost perfect. Also, in this limit both rapidities 
approach the value of $Y$, i.e. the YK and LSPS systems coincide with the 
LCMS. The reason for this is that for large $K_\perp$ the Boltzmann term in
the emission function (\ref{2.1}) becomes sharply peaked around the point
$x$ where the fluid velocity agrees with the pair velocity; the geometric 
terms in the emission function are much smoother and can be neglected. 
The relevant term is the first term in Eq.~(\ref{2.5}), and
thus the relevant variable is the transverse mass $M_\perp$. Hence this 
kinematic region starts for kaons at smaller values of $K_\perp$ than 
for pions (see Fig.~\ref{F1}).

For small values  of $M_\perp$, the difference $Y_{_{\rm YK}} 
- Y_{_{\rm LSPS}}$ increases, and both begin to lag behind the 
LCMS rapidity $Y$. Still, the YK frame is closer to the LCMS 
than is the LSPS.

The fact that the YK and LSPS systems track each other so closely implies 
that the linear rise of the YK rapidity with the pair rapidity $Y$ 
reflects nothing but a similar rise of the LSPS rapidity with $Y$.
As argued above, the latter is a direct indication for longitudinal expansion 
of the source. However, it should be noted that this expansion need not
necessarily be of hydrodynamic nature. The same feature would be generated
by a source consisting of free-streaming pions and resonances which were 
created at an initial proper time $\tau_{\rm form}$ through a 
boost-invariant production mechanism \cite{BJ83}, suffering no further 
re-scattering. It is easily seen that the strict correlations between 
coordinates and momenta in a free-streaming gas again lead to a linear
dependence of the ``source rapidity'' $Y_{_{\rm YK}}$ on the pair rapidity,
with $M_\perp$-independent unit slope. In fact, it is possible to simulate
this situation with the emission function (\ref{2.1},\ref{2.3}) by setting 
$T$ and $\eta_f$ to zero, i.e. by eliminating the thermal smearing of 
the momenta and the transverse collective flow. (Of course, this would also
result in vanishing YKP radius parameters, because pions of fixed rapidity
$Y$ can come from only a single point in the source.)

A linear rise of $Y_{_{\rm YK}}$ with $Y$ (with approximately unit slope) was
recently observed by the GIBS collaboration in Dubna~\cite{GIBS96} by analysing
pion correlations from Mg+Mg collisions at 4.4~$A$~GeV/$c$. They interpreted 
their result as evidence for rapid longitudinal expansion of the source.
The data were averaged over the transverse pair momentum $K_\perp$.
The data sample was taken with a ``central'' trigger, but since $^{24}$Mg 
is a rather small nucleus with a large surface to volume ratio it is not
clear what fraction of the participating nucleons were stopped to become
part of a thermalized fireball. The strong linear increase of $Y_{_{\rm YK}}$
with $Y$ could thus also reflect to some part the free-streaming expansion
of the pion sources created in the periphery of the nuclear reaction.
Preliminary results of the NA49 collaboration at CERN for Pb+Pb collisions 
at 158~$A$~GeV/$c$ also show a rise of $Y_{_{\rm YK}}$ with
the pair rapidity $Y$ \cite{NA49}. Here, however, the $Y$-dependence
of $Y_{_{\rm YK}}$ does not appear to be quite linear, and its slope is 
less then 1. With all due caution with regard to the still preliminary nature 
of these results, this may indicate genuine hydrodynamic longitudinal 
expansion at a somewhat slower rate than resulting from our longitudinal 
scaling profile. NA49 have also looked separately at a subsample of 
pairs with $K_\perp > 300$ MeV/$c$, showing that for them the YK rapidity 
appears to rise more rapidly with $Y$ than in the $K_\perp$-averaged sample, 
in agreement with theoretical expectations for a thermalized source 
(see. Fig.~\ref{F1}a).

The $M_\perp$-dependence of $Y_{_{\rm YK}}$ can be understood along the same
lines. Fig.~\ref{F1}b shows that $Y_{_{\rm YK}}(Y)$ approximates $Y$
better with increasing transverse mass. In the limit $M_\perp \to \infty$ 
the bosons can be emitted only from the source element which moves exactly
with the same rapidity, hence $Y_{_{\rm YK}} \to Y$. In the opposite limit
$M_\perp \to 0$ the Boltzmann factor in (\ref{2.1}) becomes a smooth 
function of $x$, and the emission function is dominated by the 
Gaussian geometric terms. In this limit one thus expects the YK rapidity to 
approach the value $Y_{_{\rm YK}} = \eta_0$. The numerical results of 
Fig.~\ref{F1}b show that the value of the pion mass is already large enough 
for the Boltzmann part of the emission function to become important. As a
result the $M_\perp$-dependence of $Y_{_{\rm YK}}$ is weak in the entire 
range which can be covered by pions, and even weaker for kaons.

\subsection{Correction terms}
\label{sec4b}

In this subsection, we study quantitatively the correction terms
of Eqs.~(\ref{1.19}) and (\ref{3.1}) which may compromise the approximation 
(\ref{1.18}) for $v$ and the simple interpretation of $R_\parallel$ and 
$R_0$ as longitudinal and temporal widths of the emission function.
Since the geometric interpretation of $R_\parallel$ and $R_0$ refers
to the YK frame, our analysis will also be done in this frame.

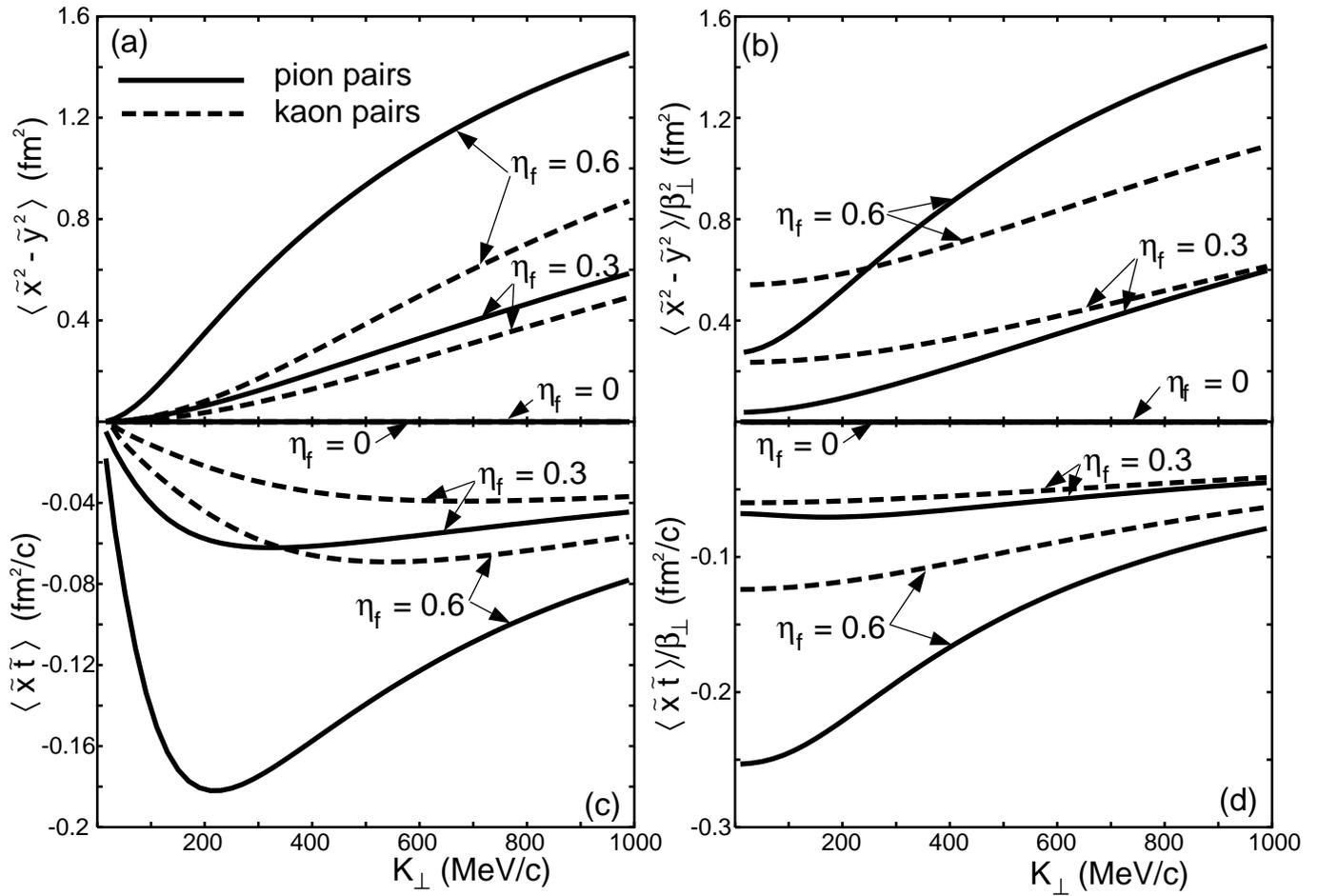
\begin{figure}
\caption{
  The correction terms $\langle \tilde x^2 - \tilde y^2 \rangle$ (a) and 
  $\langle \tilde x \tilde t \rangle$ (c), for pairs with $Y=0$ in the CMS,
  as functions of $K_\perp$ for different values of $\eta_f$. The third
  correction term $\langle \tilde x \tilde z \rangle$ vanishes at $Y=0$.
  Solid (dashed) lines refer to pion (kaon) pairs. Figures (b) and (d) show
  the same quantities, but scaled by the appropriate inverse powers of
  $\beta_\perp$ (see text).
}
\label{F2}
\end{figure}

Let us first focus on the central rapidity region $Y = \eta_0$. Then 
$Y_{_{\rm YK}} = Y_{_{\rm LSPS}}=Y$, i.e., the four reference frames 
listed in Sec.~\ref{sec4a} coincide. Furthermore, the source is symmetric 
in the longitudinal direction and thus $\langle \tilde x \tilde z \rangle 
= \langle \tilde z \tilde t \rangle \equiv 0$. The only non-vanishing 
corrections thus arise from the terms $\langle \tilde x \tilde t \rangle$ and
$\langle {\tilde x}^2 - {\tilde y}^2 \rangle$. In Fig.~\ref{F2} they are
plotted as a function of $K_\perp$ for different values of the scaling 
parameter ${\eta}_f$ for the transverse flow. Without transverse flow (i.e. 
for ${\eta}_f = 0$) the source is azimuthally symmetric (see Eq.~(\ref{2.5}))
which implies $\langle {\tilde x}^2 - {\tilde y}^2 \rangle =0$. Also, the 
source is reflection symmetric in the out-direction, $\langle \tilde x 
\tilde t \rangle = 0$. For non-zero transverse flow the correction
terms are generally non-zero, and they grow with increasing $\eta_f$.
Note that, for fixed $\eta_f$ and $K_\perp$, the correction terms are 
considerably smaller for kaons than for pions. This can be understood
as follows: as discussed after Eq.~(\ref{2.5}), for $\eta_f=0$ the source 
depends only on $M_\perp$, and thus even for non-zero transverse flow 
everything to zeroth order still scales with $M_\perp$. We will 
show below that for expanding sources the regions of homogeneity, 
which effectively contribute to the correlation function, are 
generically decreasing functions of $M_\perp$. This is thus also true for 
the correction terms. Since at fixed $K_\perp$ the value of $M_\perp$
is larger for kaons than for pions, the corresponding correction terms 
are smaller in absolute terms (although not necessarily relative to the
leading contributions).

The $K_{\perp}$-dependence of the correction terms at non-zero transverse 
flow $\eta_f \ne 0$ can also be easily understood. The rise of $\langle 
\tilde x^2 - \tilde y^2 \rangle$ for increasing transverse momentum 
(Fig.~\ref{F2}a) is due to the azimuthal symmetry breaking by the second 
term of Eq.~(\ref{2.5}) which increases both with $K_{\perp}$ and $\eta_f$. 
It agrees with the findings of Ref.~\cite{WSH96} but, as pointed out
in \cite{HV96}, contradicts the behaviour seen by Pratt in the first of
Refs.~\cite{P84} for an infinitesimally thin spherically expanding
shell where $\langle \tilde x^2 - \tilde y^2 \rangle$ decreases
with increasing $K_\perp$. A similar behaviour is seen \cite{SH97} 
in hydrodynamical simulations where freeze-out occurs along an 
infinitesimally thin freeze-out hypersurface; there also $\langle \tilde x^2 
- \tilde y^2 \rangle$ first decreases very rapidly with increasing 
$K_\perp$, then saturates and slightly increases again without,
however, ever turning positive. The strong decrease of $\langle \tilde x^2 
- \tilde y^2 \rangle$ with $K_\perp$ appears to be an artefact of the 
idealization of an infinitesimally thin expanding shell; in \cite{TH97} 
it was shown to be much less visible for opaque sources with a finite 
thickness of the emitting surface layer, returning to the here observed 
rise already for a rather modest surface thickness.

Different from $\langle \tilde x^2 - \tilde y^2 \rangle$, the variance
$\langle \tilde x \tilde t \rangle$ reaches an extremum and then decreases 
again for very large $K_\perp$ (Fig.~\ref{F2}c). This results from an 
interplay between the increasing breaking of the $\tilde x \to - 
\tilde x$ reflection symmetry, which tends to increase the value 
for $\langle \tilde x \tilde t \rangle$, and a decreasing homogeneity 
length in space-time rapidity $\eta$ which affects the $t=\tau\cosh\eta$ 
part of this variance.

In the model-independent expressions (\ref{1.19}) and (\ref{3.1}) for the 
YKP parameters, the correction terms discussed above are divided by 
${\beta}_\perp$ and ${\beta}_\perp^2$, respectively. From Appendix~\ref{app} 
we know that the ratios remain finite in the limit $\beta_\perp \to 0$.
Still, the corrections to the YKP parameters could become sizeable, 
depending on how slowly the variances $\langle \tilde x \tilde t 
\rangle$ and $\langle \tilde x^2 - \tilde y^2 \rangle$ vanish in this limit.
In Fig.~\ref{F2}b,d we show that the correction terms actually remain 
small even after dividing them by the appropriate powers of $\beta_\perp$. 
Thus, at least at mid-rapidity, the leading order approximations 
(\ref{1.19}b,c) are seen to be generally very good. The largest 
correction comes from the difference $\langle \tilde x^2 - \tilde y^2 
\rangle / \beta_{\perp}^2$ in Eq.~(\ref{1.19c}). A more detailed discussion
of its effects on $R_0$ will follow in the next subsection.

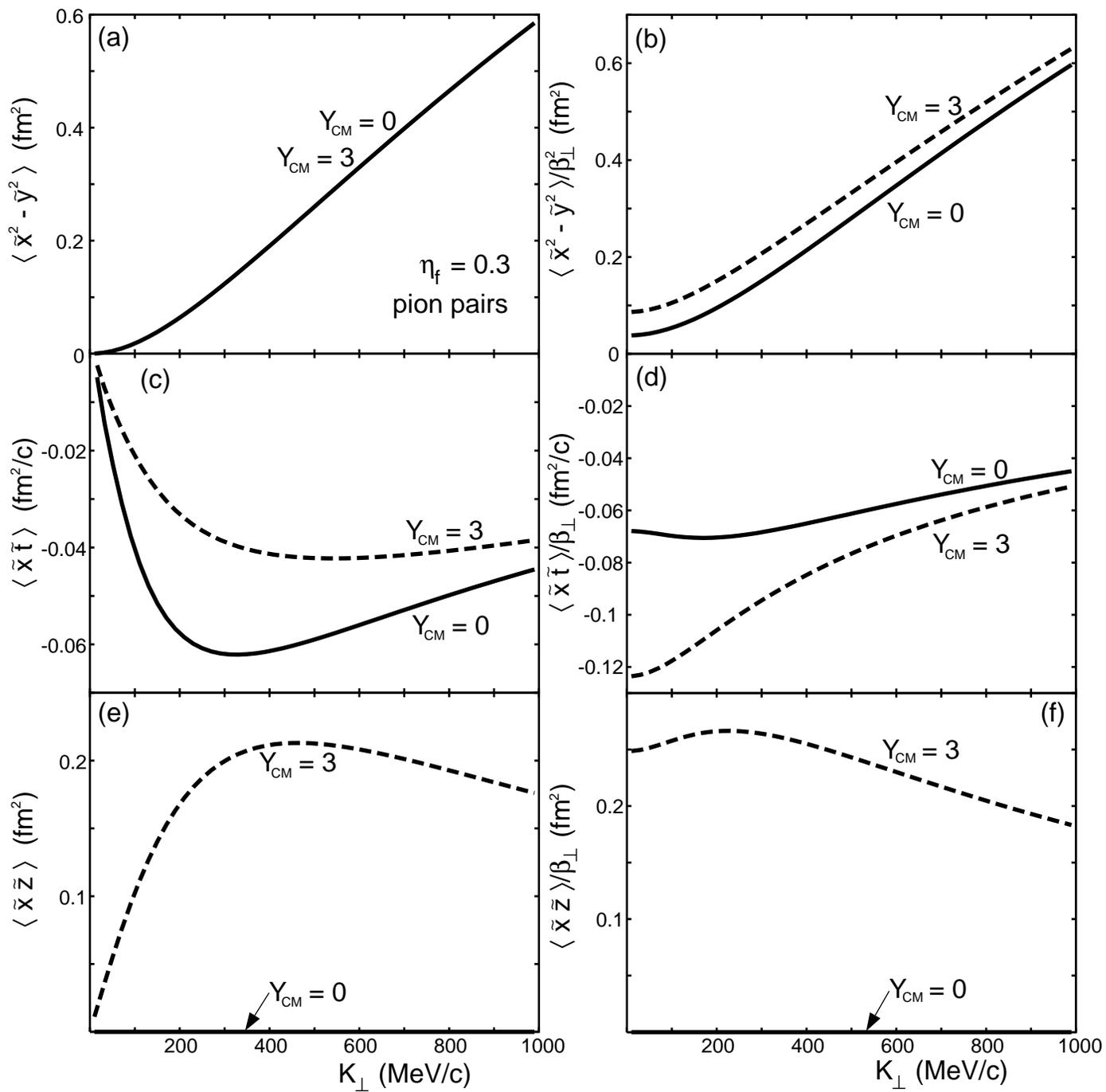
\begin{figure}
\caption{ 
  Same as Fig.~\protect\ref{F2}, but now for a fixed transverse flow
  $\eta_f=0.3$ and pions only, but for two different pair rapidities, 
  $Y_{_{\rm CM}}=0$ (solid) and $Y_{_{\rm CM}}=3$ (dashed). The curves
  were evaluated in the YK frame.
}
\label{F3}
\end{figure}

We now proceed to a discussion of the correction terms for $Y \ne \eta_0$.
We define $Y_{_{\rm CM}}=Y-\eta_0$ as the rapidity of the pair in the CMS.
In Fig.~\ref{F3} we compare the correction terms for $Y_{_{\rm CM}} = 3$ to
those for $Y_{_{\rm CM}} = 0$. Of course, at  $Y_{_{\rm CM}} = 3$ the YK 
frame no longer coincides with the CMS, see Sec.~\ref{sec4a}. Since 
the transverse variances are not affected by longitudinal boosts nor 
do they depend on $Y$ (see Appendix~\ref{app}), $\langle \tilde x^2 - 
\tilde y^2 \rangle$ does not change with the pair rapidity. However, at
a given value of $K_\perp$ the transverse pair velocity ${\beta}_\perp = 
K_\perp/E_K$ becomes smaller, since $E_K$ increases with the pair rapidity
$Y_{_{\rm CM}}$. The correction term $\langle \tilde x^2 - \tilde y^2 \rangle
/ \beta_\perp^2$ thus increases with $Y_{_{\rm CM}}$, especially at low 
$K_\perp$. A similar effect is seen in the plots for $\langle \tilde x 
\tilde t \rangle/\beta_\perp$. While $\langle \tilde x \tilde t \rangle$ 
decreases with increasing $Y_{_{\rm CM}}$, the ratio with $\beta_\perp$
actually increases by about a factor 2 at small $K_\perp$. Since in the 
forward rapidity region the source becomes non-symmetric under 
reflection $z \to -z$, $\langle \tilde x \tilde z \rangle$ and 
$\langle \tilde x \tilde z \rangle /\beta_\perp$ are non-zero, but small.

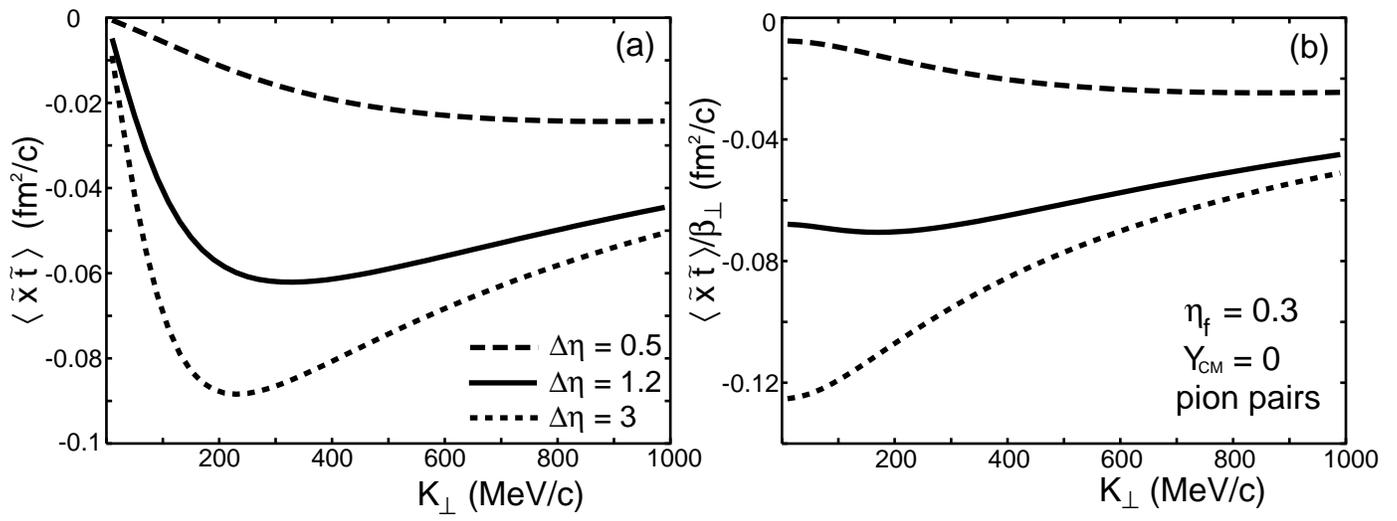
\begin{figure}
\caption{ 
  Dependence of the correction term 
  ${\langle{ \tilde{x}\tilde{t} }\rangle}(K_{\perp})$ on the
  longitudinal width parameter $\Delta \eta$, for pion pairs with
  $Y_{_{\rm CM}} = 0$.
}
\label{F4}
\end{figure}

We conclude this subsection with a discussion of the sensitivity of the
correction terms to the longitudinal extension of the source, which 
is parametrised by $\Delta \eta$. We again put $Y_{_{\rm CM}}=0$, so
that $\langle \tilde x \tilde z \rangle = 0$. From the explicit expressions
given in Appendix~\ref{app} it is clear that the transverse variances 
$\langle \tilde x^2 - \tilde y^2 \rangle$ are independent of $\Delta\eta$. 
Therefore only the size of $\langle \tilde x \tilde t \rangle$
changes. In Fig.~\ref{F4} we show its $K_\perp$-dependence for different
values of $\Delta \eta$. Due to our assumption of freeze-out along
a hyperbola of constant proper time $\tau_0$ (smeared by an amount 
$\Delta\tau$), an increase of $\Delta \eta$ causes a larger effective 
extension of the source both in the $z$ and in the time direction. This 
becomes especially important at small $K_\perp$ where the geometrical 
factors in the emission function (\ref{2.1}) play an important role. This 
explains the relatively large $\Delta\eta$-dependence of $\langle \tilde x 
\tilde t \rangle$ at small $K_\perp$ and the weaker dependence at large
$K_\perp$ where the source distribution is dominated by the Boltzmann
term which does not depend on $\Delta \eta$. 

\subsection{Quantitative numerical study of YKP radius parameters}
\label{sec4c}

In this subsection we combine the results from the previous subsection 
for the correction terms with the leading contributions in order
to arrive at a quantitative understanding of the longitudinal and 
temporal YKP radius parameters $R_\parallel$ and $R_0$, and in particular 
of their dependence on the pair momentum ${\bf K}$. 

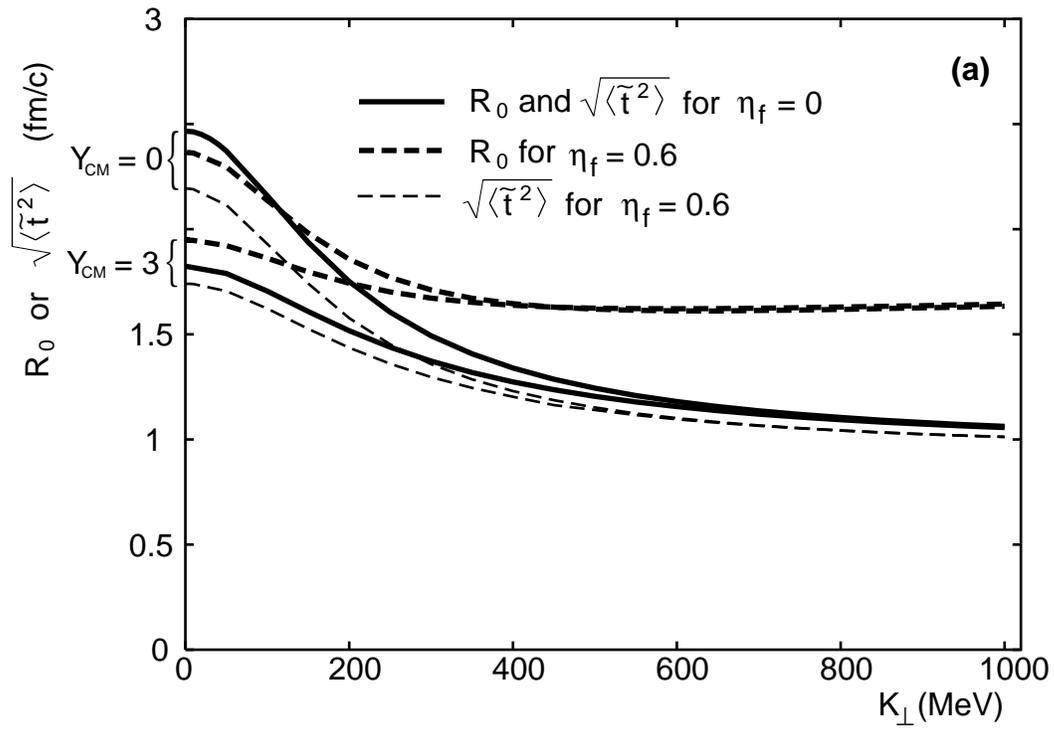
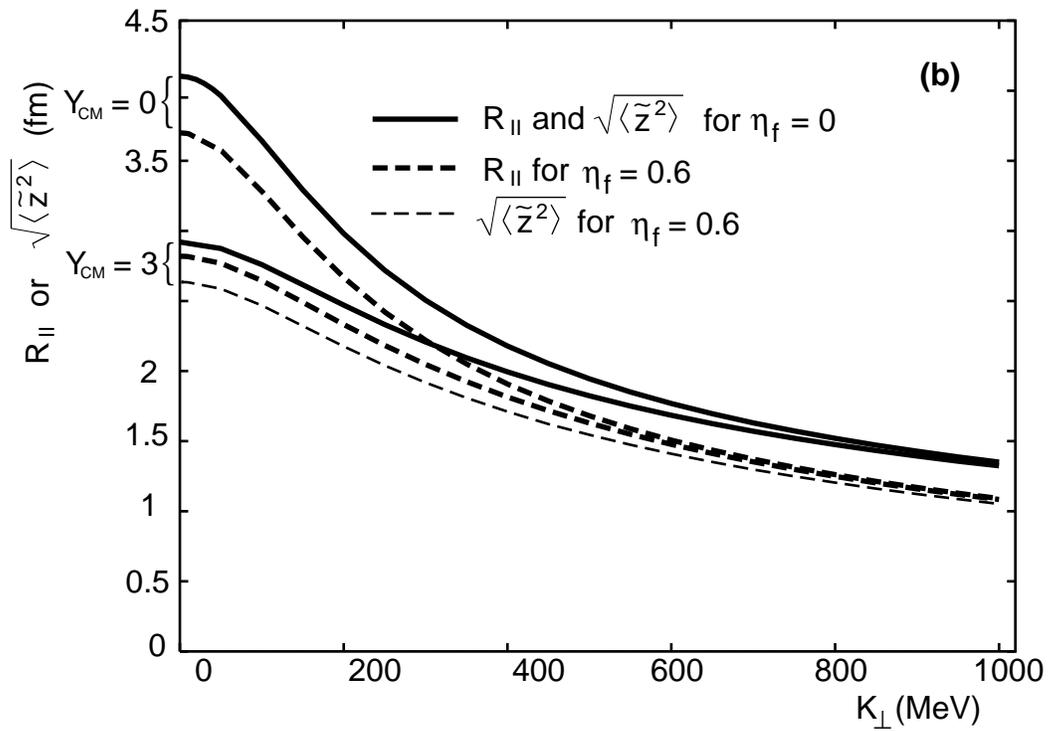
\begin{figure}
\caption{
 (a) $R_0$ and $\protect \sqrt{\langle \tilde t^2 \rangle}$
     as functions of $\protect K_\perp$ for
     $\eta_f = 0$ and $\eta_f = 0.6$,
     for pion pairs with rapidity $Y_{_{\rm CM}}=0$ and $Y_{_{\rm CM}}=3$.
     The lifetime $\protect \sqrt{\langle \tilde t^2 \rangle}$ is evaluated 
     in the YK rest frame.
 (b) Same as (a), but for $R_\parallel$ and the
     longitudinal length of homogeneity $\protect \sqrt{\langle \tilde
     z^2 \rangle}$ in the YK rest frame. For $Y_{_{\rm CM}}=0$, 
     $R_\parallel$ and $\protect \sqrt{\langle \tilde z^2 \rangle}$ 
     agree exactly because $\beta_l=0$ in the YK frame.
}
\label{F5}
\end{figure}

In Fig.~\ref{F5} we show $R_0$ and $R_{\parallel}$ together with
their approximations $\sqrt{\langle\tilde t^2\rangle}$,
$\sqrt{\langle\tilde z^2\rangle}$, for pion pairs with rapidity 
$Y_{_{\rm CM}} = 0$ and $Y_{_{\rm CM}} = 3$, respectively, as functions of
$K_{\perp}$. For vanishing transverse flow both approximations are seen 
to be exact, in agreement with the discussion from the previous subsection.
For non-zero transverse flow the approximation $R_\parallel \approx 
\sqrt{\langle \tilde z^2 \rangle}$ remains exact for pairs with 
$Y_{_{\rm CM}}=0$. The reason is that for such pairs the YK rapidity
relative to the CMS is zero, and thus the longitudinal velocity $\beta_l$ 
(which multiplies the correction terms in (\ref{1.19b})) of the pair
in the YK frame vanishes. From Fig.~\ref{F5}b one sees, however, that
also for forward rapidity pairs at $Y_{_{\rm CM}}=3$ the correction 
terms stay below 10\% at all values of $K_{\perp}$.

The situation is not quite as good for $R_0$. Here one sees apparently
strong differences between $R_0$ and $\sqrt{\langle \tilde t^2 \rangle}$
as soon as the transverse flow is switched on. From Fig.~\ref{F2}
it is clear that the (in our case positive) correction term 
$\langle \tilde x^2 - \tilde y^2 
\rangle$ is the culprit and dominates the difference. For a transverse 
flow of ${\eta}_f = 0.6$ as shown in Fig.~\ref{F5}a this term becomes 
(in the experimentally accessible $K_\perp$ range) larger than 
1~(fm/$c$)$^2$ and thus comparable to the leading term $\sqrt{\langle 
\tilde t^2 \rangle}$. However, the 
numerical results shown in this Figure actually correspond to a rather
extreme situation. First, the assumed transverse flow rapidity 
$\eta_f=0.6$ is large; the heavy ion data at AGS and CERN 
energies for Si- and S-induced reactions seem to require smaller values
\cite{SH92,Stachel94,Herrmann96}. Second, the difference between $R_0$
and $\sqrt{\langle \tilde t^2 \rangle}$ is small at low transverse
momenta and becomes large only at large $K_\perp$; in that range the
leading term $\sqrt{\langle \tilde t^2 \rangle}$ is essentially given
by the source parameter $\Delta\tau$ (see discussion below) which in
Pb+Pb and Au+Au collisions \cite{NA49,CN96} (where $\eta_f$ may be larger 
than for the smaller systems analyzed so far) seems to be bigger than the 
1 fm/$c$ assumed here\footnote{If the source, unlike ours, is opaque, 
i.e. if the particle emission is strongly surface dominated, ${\langle{ 
\tilde{x}^2 - \tilde{y}^2 }\rangle}$ tends to be negative 
\cite{HV96,TH97,SH97}. In this case, the deviation of $R_0$ from 
$\sqrt{ {\langle{\tilde{t}^2}\rangle}}$ will have the opposite sign 
\cite{HV96,SH97}, and $R_0^2$ usually even turns negative for small values 
of $K_\perp$. For a detailed study we refer to \cite{TH97}.}.

Fig.~\ref{F5}a shows that the effective source lifetime $\Delta t =
\sqrt{\langle \tilde t^2 \rangle}$ is a strong function of the pair 
momentum ${\bf K}$: it is largest at small rapidity $Y_{_{\rm CM}}$ 
and transverse momentum $K_\perp$ and decreases with increasing 
$Y_{_{\rm CM}}$ and/or $K_\perp$. Its asymptotic value for large 
${\bf K}$ in the CMS is, not unexpectedly, given by the
variance (\ref{A11}) of the proper time distribution of our source 
(\ref{2.1}). But why is it larger for pairs with smaller momenta ${\bf K}$?

From Fig.~\ref{F5}b it is clear that the longitudinal region
of homogeneity $R_\parallel$ is a decreasing function of the pair
momentum ${\bf K}$. The reason for this is the same as the similar
decrease of $R_l$ in the Cartesian fit and well understood 
\cite{MS88} as a consequence of the strong longitudinal 
expansion of the source. This expansion introduces a longitudinal
velocity gradient, and the longitudinal length of homogeneity is given by 
the inverse of this gradient multiplied by a ``thermal smearing factor''
\cite{CSH95b}. The latter reflects the statistical distribution of the 
particle momenta around the local source fluid velocity, and for a thermal
distribution the spatial region over which this thermal smearing is 
effective decreases with increasing pair momentum. This causes the 
shrinking of the longitudinal homogeneity length with ${\bf K}$.

Since for different pair momenta $R_0$ measures the source lifetime
in different YK reference frames, the freeze-out ``hypersurface'' will
in general appear to have different shapes for pairs with different momenta.
Only in our model, where freeze-out occurs at fixed proper time $\tau_0$
(up to a Gaussian smearing with width $\Delta\tau$), is it frame-independent.
It is thus generally unavoidable (and here, of course, true in any frame)
that freeze-out at different points $z$ in the source will occur at different
times $t$ in the YK frame. Since a $z$-region of size $R_\parallel$ 
contributes to the correlation function, $R_\parallel$ determines how large 
a domain of this freeze-out surface (and thus how large an interval of
freeze-out times in the YK frame) is sampled by the correlator. This
interval of freeze-out times combines with the intrinsic Gaussian width 
$\Delta\tau$ to yield the total effective duration of particle emission.
It will be largest at small pair momenta where the homogeneity region 
$R_\parallel$ is biggest, and will reduce to just the variance of the
Gaussian proper time distribution at large pair momenta where the
longitudinal (and transverse) homogeneity regions shrink to zero.

Another interesting feature of Fig.~\ref{F5} is that at large $K_\perp$ 
both $R_\parallel$ and $R_0$ are independent of the pair rapidity $Y$. 
This is a consequence of our boost-invariant longitudinal velocity profile
and need not remain true for systems with different longitudinal expansion.
As argued before, at large $M_\perp$ the space-time shape of the source
is dominated by the Boltzmann term and becomes insensitive to the Gaussian 
geometric factors. The HBT radii thus only see the local velocity
gradients which in our case are invariant under longitudinal boosts.
At large $M_\perp$ pion pairs with different rapidities $Y$ thus all
see the same local source structure, and the YKP radii become $Y$-independent.

We close this subsection with a discussion of the dependence of 
$R_\parallel$ and $R_0$ on the other source parameters. Since the rapidity
dependence does not change qualitatively from what has already been
discussed, we concentrate on zero rapidity pion pairs, $Y_{_{\rm CM}}=0$.
For the transverse flow we choose a non-zero, but moderate value 
of $\eta_f=0.3$.

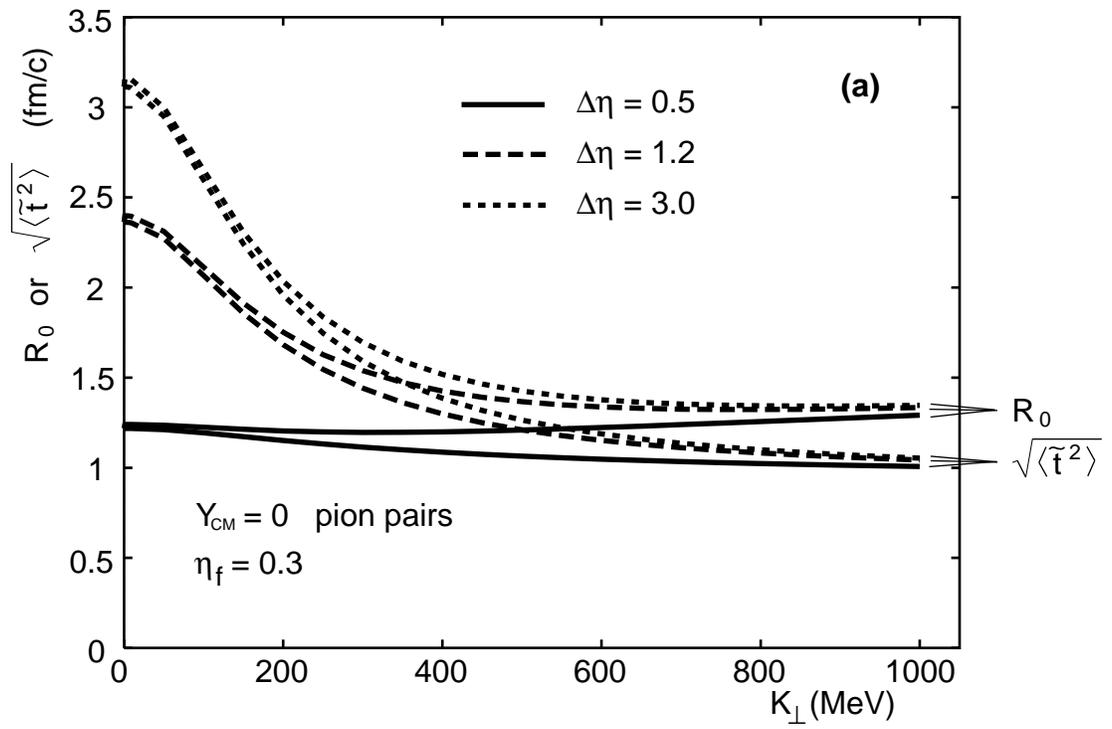
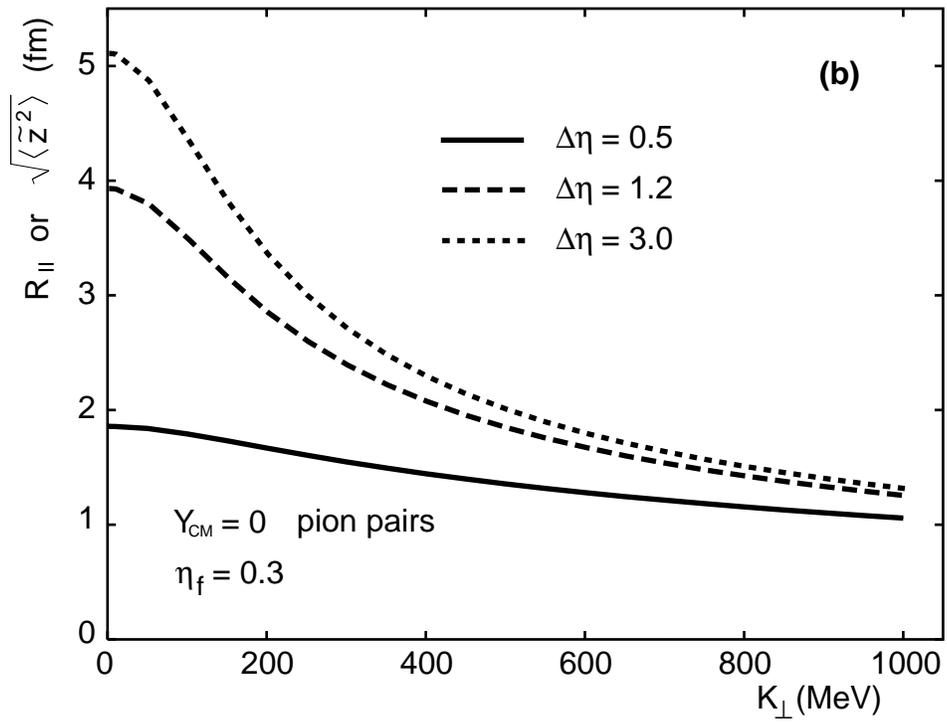
\begin{figure}
\caption{
 (a) $R_0$ and $\protect \sqrt{\langle \tilde t^2 \rangle}$
     as functions of $\protect K_\perp$, for different
     longitudinal gaussian widths $\Delta\eta$. The diagrams are
     for pion pairs with $Y_{_{\rm CM}}=0$ and for transverse 
     flow $\eta_f = 0.3$. The lifetime $\protect \sqrt{\langle 
     \tilde t^2 \rangle}$ is evaluated in the YK rest
     frame which coincides here with the CMS and the LCMS.
 (b) Same as (a), but for $R_\parallel$. Since $Y_{_{\rm CM}}=0$,
     $R_\parallel$ and the longitudinal length of homogeneity
     in the YK frame $\protect \sqrt{\langle \tilde
     z^2 \rangle}$ agree exactly.
}
\label{F6}
\end{figure}

In Fig.~\ref{F6} we show the dependence on the longitudinal size of the
source which is parametrised by $\Delta\eta$. One sees that at large 
transverse momenta neither $R_\parallel$ nor $R_0$ and $\sqrt{\langle 
\tilde t^2 \rangle}$ are affected by the width $\Delta \eta$ of the 
Gaussian geometric factor, in line with the arguments above. At small
transverse momenta, both radii increase monotonically with $\Delta\eta$.
This means that, at low $K_\perp$, $R_\parallel$ becomes sensitive to the
global longitudinal geometry of the source and no longer only reflects
the local longitudinal velocity gradients. Fig.~\ref{F6}a is very 
interesting: for small $\Delta\eta$ the longitudinal length of homogeneity
is limited by the longitudinal geometry, and $R_0$ never has a chance
to probe a large region of the proper time freeze-out surface. Hence
$\Delta t$ is limited to the variance of the proper time distribution 
$T(\tau)$ in the source (see (\ref{A2},\ref{A11})), independent of 
$K_\perp$. As $\Delta \eta$ increases, at small $K_\perp$ the longitudinal
length of homogeneity increases too, and $R_0$ receives an additional
contribution from the time variation (in the YK rest frame) along
the freeze-out surface inside a longitudinal region of size $R_\parallel$.
Thus the rise of the effective source lifetime $\Delta t$ at small
$K_\perp$ is an indirect measure for the longitudinal geometric size of 
the source. Unfortunately, the detailed quantitative dependence
is model-dependent.

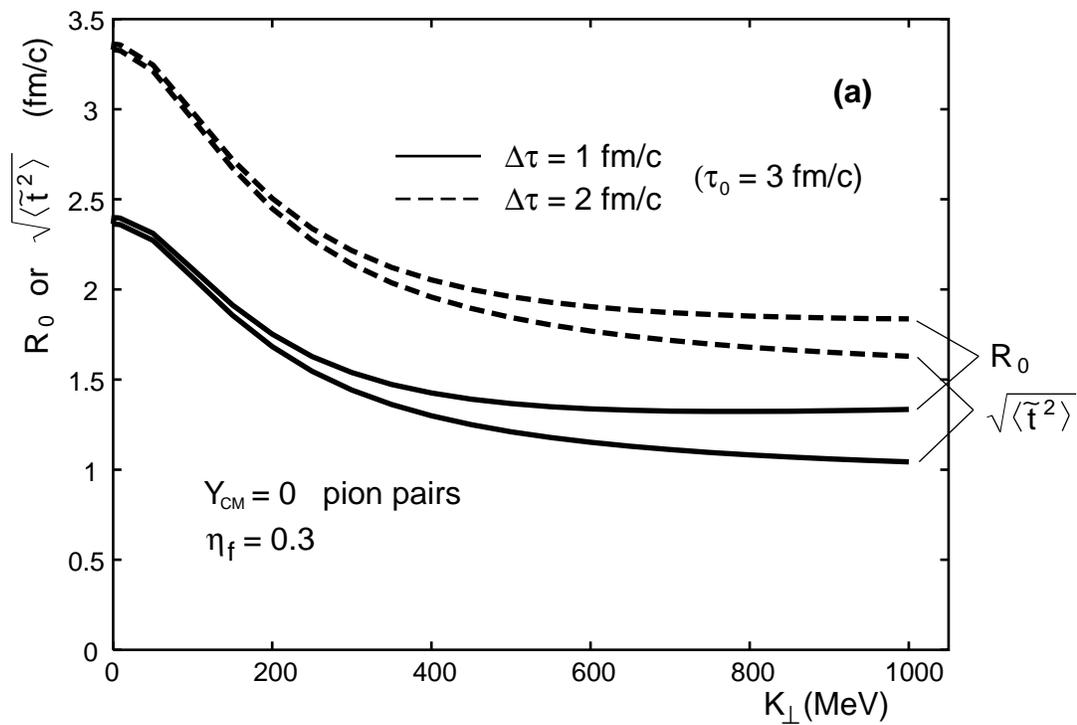
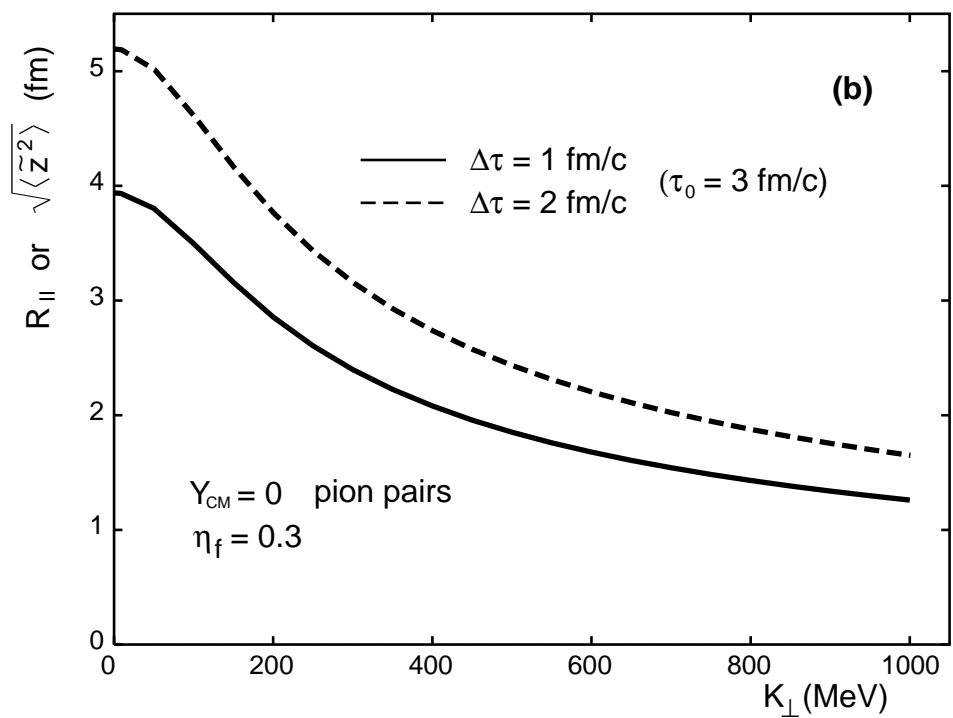
\begin{figure}
\caption{Same as Fig.~\protect\ref{F6}, but for different widths
     $\Delta\tau$ of the proper time distribution in the source.
}
\label{F7}
\end{figure}

Fig.~\ref{F7} shows what happens when the width of the proper time
distribution $T(\tau)$ (Eq.~(\ref{A2})) is changed. Increasing 
$\Delta\tau$ by 1 fm/$c$, $R_0$ also increases by about 1 fm/$c$ 
(slightly less at large $K_\perp$\footnote{Actually, the proper way 
of looking at this increase is by studying the variance of the function 
$T(\tau)$: according to Eq.~(\ref{A11}) it increases from 0.89 fm/$c$ 
for $\Delta\tau=1$ fm/$c$ to 1.49 fm/$c$ for $\Delta\tau=2$ fm/$c$.}), 
while $R_\parallel$ 
increases more at small $K_\perp$ and less at large $K_\perp$. The 
increase of $R_\parallel$ is due to the decrease of the longitudinal
velocity gradient (which for a boost-invariant profile is $1/\tau$)
with $\tau$. As the time distribution $T(\tau)$ becomes wider, larger
proper times are probed by the emitted pions resulting in larger 
longitudinal homogeneity regions.

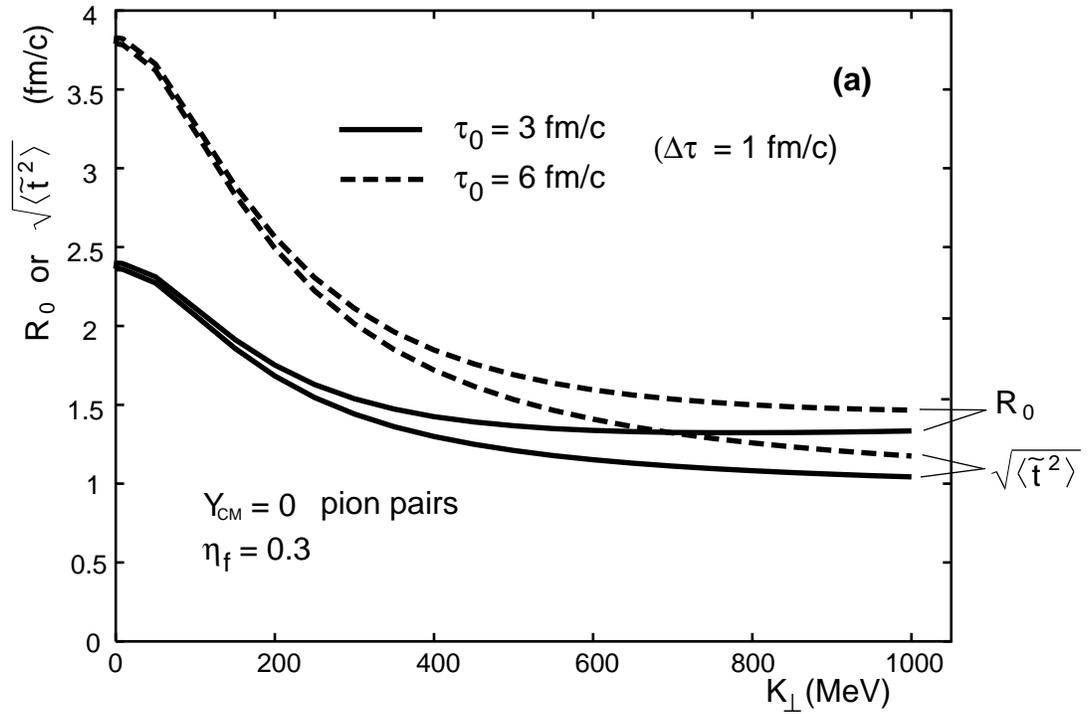
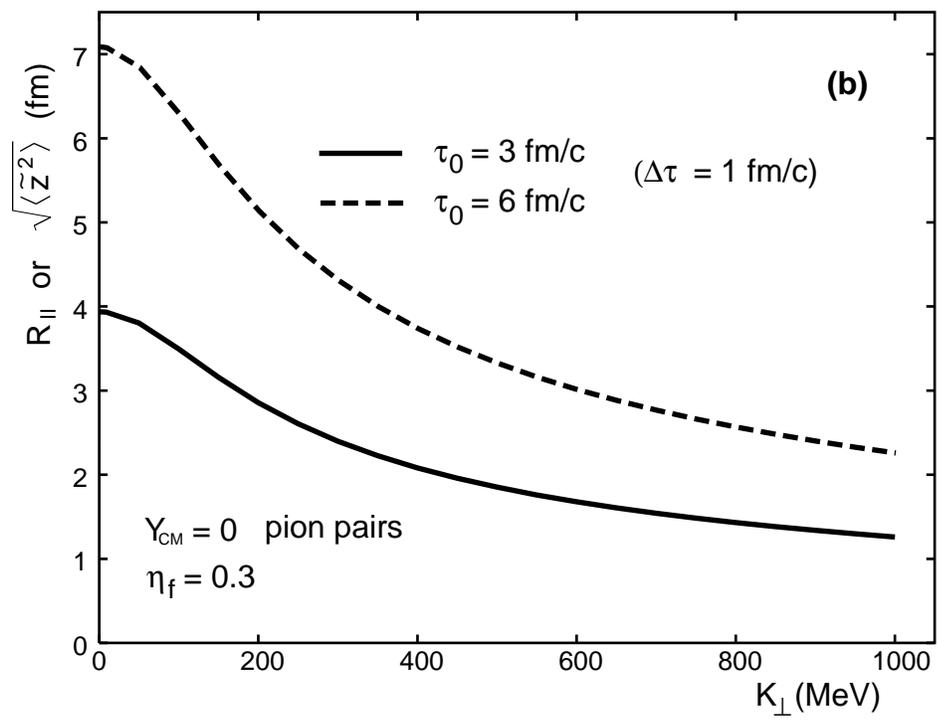
\begin{figure}
\caption{Same as Fig.~\protect\ref{F6}, but for different average
     freeze-out times $\tau_0$.
}
\label{F8}
\end{figure}

Changing the average freeze-out time $\tau_0$ rather than its spread
$\Delta\tau$ has qualitatively similar consequences (see Fig.~\ref{F8}),
only that $R_0$ at sufficiently large $K_\perp$ again reduces
to the same small variance of the time distribution $T(\tau)$. 
Note that, at small $K_\perp$, $R_0$ increases both with increasing 
$\Delta \eta$ and increasing $\Delta\tau$; this supports our claim 
that it is ``sensitive'' to the total longitudinal extension of the source
$\Delta z \simeq 2 \tau_0\sinh\Delta\eta$. However, the relation is not
linear (in particular in our numerical results a doubling of 
$\Delta\eta$ is seen to have less effect than a
doubling of $\tau_0$), making it hard to use in practice. It is obvious
that the longitudinal velocity gradient (which decreases by a factor 2
when $\tau_0$ is doubled) has a stronger influence on $R_0$ and $R_\parallel$
than the geometrical width in $\eta$. 

\subsection{Transverse flow, $M_\perp$-scaling, and kaon interferometry}
\label{sec4d}

In this subsection we compare pion and kaon correlation functions.
We discuss the $M_\perp$-scaling of the YKP radius parameters and its
breaking by transverse collective flow. At the end of the subsection we 
formulate a program how to extract transverse flow from the 
$M_\perp$-dependence of the YKP radius parameters.

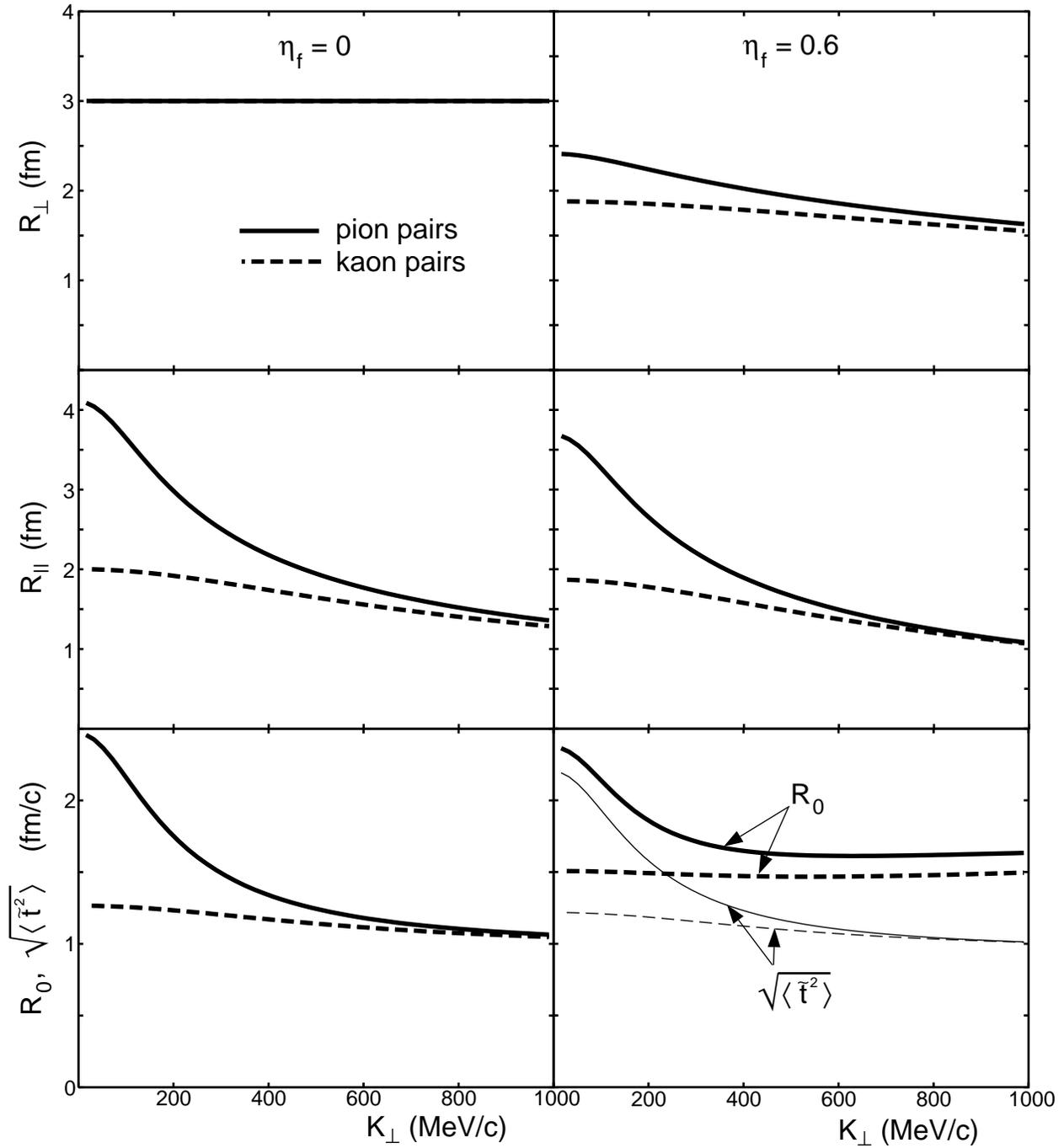
\begin{figure}
\caption{
     The YKP radii $R_\perp$ (top row), $R_\parallel$ (middle row),
     and $R_0$ (bottom row), for $Y_{_{\rm CM}}=0$ pion (solid) and 
     kaon (dashed) pairs, as functions of $K_\perp$. 
     Left column: no transverse flow. Right column: transverse flow 
     $\eta_f=0.6$. In the lower right panel we also show 
     the effective lifetime in the YK frame $\protect \sqrt{\langle \tilde t^2 
     \rangle}$ for comparison. For more discussion see text. 
}
\label{F9}
\end{figure}

In Fig.~\ref{F9} we compare, for central rapidity pairs $Y_{_{\rm CM}}=0$, 
the three YKP radius parameters for pion and kaon pairs, as functions of
$K_\perp$. The left column shows a source without transverse expansion,
in the right column the transverse flow rapidity was set to $\eta_f=0.6$.
The onset of transverse flow has two qualitative effects: {\em (i)} the 
transverse radius acquires a $K_\perp$-dependence \cite{WSH96}, and 
{\em (ii)} $R_0$ and $\sqrt{\langle \tilde t^2 \rangle}$ begin to deviate
from each other, as discussed in the previous subsection. (The 
equality $R_\parallel^2 = \langle \tilde z^2 \rangle$ remains exact
because we are studying pion pairs at $Y_{_{\rm CM}}=0$.) The effects of 
flow on $R_\parallel$ and $\sqrt{\langle \tilde t^2 \rangle}$ are seen to 
be weak, for both pions and kaons.

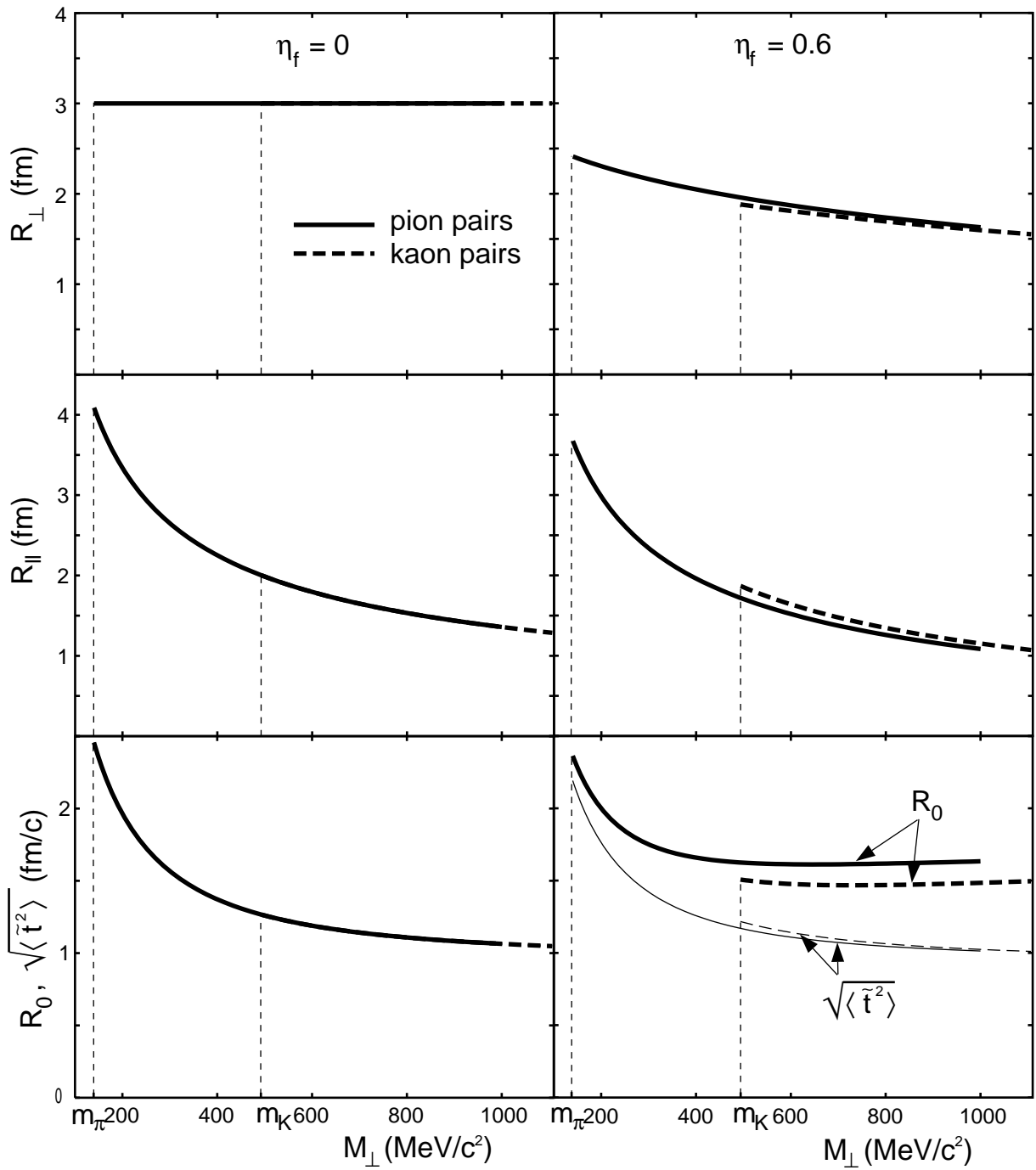
\begin{figure}
\caption{
     Same as Fig.~\protect\ref{F9}, but plotted as functions of
     $M_\perp$.\
}
\label{F10}
\end{figure}

Note also that at small $K_\perp$ the kaon radii are generically 
smaller than the pion radii, with or without transverse flow.
This is also seen in experiment \cite{NA44,E859}. However, except
for the change in the rest mass we have changed no parameters
in the emission function, so the difference must be entirely kinematic.
Indeed, it just reflects the fact that for thermalized sources like 
(\ref{2.1}) the leading dependence on the particle rest mass is
through the variable $M_\perp = \sqrt{m^2+K_\perp^2}$. As discussed after
Eq.~(\ref{2.5}) the source (\ref{2.1}) depends {\em only} on $M_\perp$
if it does not expand transversally ($\eta_t(r)=0$). In this case
the correction terms in Eqs.~(\ref{1.19}b,c) vanish exactly, and
$R_\perp^2 = \langle \tilde y^2\rangle$, $R_\parallel^2 = \langle 
\tilde z^2\rangle$, and $R_0^2 = \langle \tilde t^2\rangle$ are also 
functions of $M_\perp$ only. This is shown in the left column of 
Fig.~\ref{F10} where the YKP radii for a source without transverse flow
are plotted as functions of $M_\perp$ and seen to exactly coincide
for pions and kaons in the common $M_\perp$-range. Since in the absence
of transverse flow the Boltzmann factor in (\ref{2.1}) has no transverse
gradients (we have assumed a constant temperature $T$), the transverse
radius $R_\perp$ is $M_\perp$ independent and equal to the transverse
geometric (Gaussian) radius $R$. It was pointed out in 
Refs.~\cite{CSH95b,CL96} that transverse temperature gradients can also
cause an $M_\perp$-dependence of the transverse radius $R_\perp$; but
since the source remains in this case a function of $M_\perp$ only, the
$M_\perp$-scaling of the YKP radii persists; it can only be broken
by transverse flow.

The breaking of the $M_\perp$-scaling by transverse flow is shown in
the right column of Fig.~\ref{F10}, for $\eta_f=0.6$. It has two origins:
the emission function itself is no longer a function of $M_\perp$ only 
(see (\ref{2.5})), and the now non-vanishing correction terms in 
(\ref{1.19}b,c) depend on $\bbox{\beta}$ and thus on both $M_\perp$ 
{\em and} the rest mass $m$. It is obvious that the scaling violations 
induced by the pion-kaon mass difference are weak and require very 
accurate measurements. Furthermore, one may be worried that
resonance decay contributions to the correlation radii \cite{Bolz} 
(which we haven't discussed here) lead also to a breaking of the $M_\perp$ 
scaling, because they affect pions more than kaons, and this may make 
it difficult to isolate the transverse flow effects. We refer to the
detailed discussion of resonance decays in the context of the model 
source (\ref{2.1}) in Ref.~\cite{WH96}. That study shows, however,
that their influence on the $M_\perp$-dependence of the transverse radius
parameter $R_\perp$ is weak \cite{Heinz96,WH96}. Furthermore, resonances 
tend to increase
all three HBT radii (in particular the effective lifetime $R_0$), while
the $M_\perp$-scaling violations from transverse flow have the opposite
sign for $R_\parallel$ and $R_\perp,R_0$.

Detailed dynamical studies of the freeze-out process have shown that the
transverse gradients of the temperature across the freeze-out surface
tend to be small \cite{Peni90,Mayer95}. So the experimentally observed
$M_\perp$-dependence of the transverse radius \cite{NA35,NA44,NA49}
is presumably due to transverse flow \cite{Alber95}. It was shown in 
Ref.~\cite{WSH96} that the strength of the $M_\perp$-dependence
of $R_\perp$ increases monotonously with the strength $\eta_f$ of 
the transverse expansion. Alber \cite{Alber95} has suggested to quantify
the strength of collective flow by fitting the HBT radii to a power
law in $M_\perp$,
 \begin{equation}
 \label{power}
   R_\perp(M_\perp) \propto M_\perp^{-\alpha_\perp}\, , \qquad
   R_l(M_\perp) \propto M_\perp^{-\alpha_l}\, ,
 \end{equation}
and using the magnitude of the extracted (negative) power as a flow 
measure. He found $\alpha_l \simeq 0.5$ for $R_l$ and 
smaller values for $\alpha_\perp$, with a tendency to increase for 
larger collision systems \cite{Alber95}. He interpreted this as a signature
for strong longitudinal and weaker transverse expansion, the latter
becoming more important for larger systems.

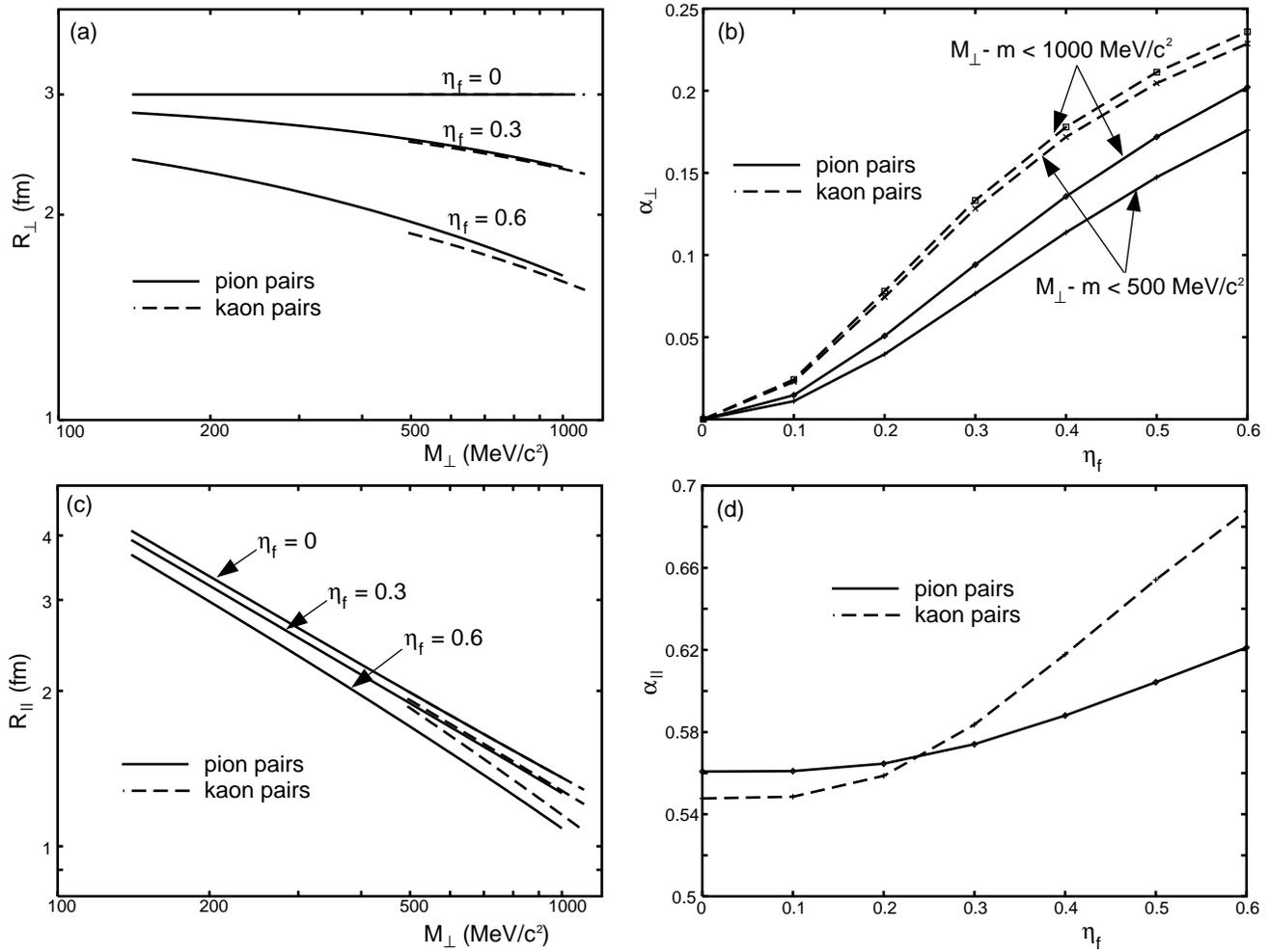
\begin{figure}
\caption{
 (a) $R_{\perp}$ as a function of $\protect M_\perp$ at $Y_{_{\rm CM}}=0$,
     for pions (solid) and kaons (dashed) and 
     different transverse flow rapidities $\eta_f$.
 (b) The scaling coefficient $\alpha_{\perp}$ defined
     by $R_{\perp} \approx M_{\perp}^{-\alpha_\perp}$ for pions (solid)
     and kaons (dashed) as a function of the transverse flow rapidity 
     $\eta_f$. The different results obtained by fitting 
     in the regions $M_{\perp} - m < 500$ MeV/$c^2$ and 
      $M_{\perp} - m < 1000$ MeV/$c^2$ are shown separately.
 (c) Same as (a), but for $R_\parallel$.
 (d) Same as (b), but for $\alpha_\parallel$.
}
\label{F11}
\end{figure}

In Fig.~\ref{F11} we study the possible conclusions from such an exercise 
when applied to the results from our model. The left column shows double
logarithmic plots for $R_\perp$ and $R_\parallel$ as functions of
$M_\perp$. Obviously the assumption of a power law dependence is well 
justified for $R_\parallel$ but somewhat marginal for $R_\perp$. 
$R_0(M_\perp)$ cannot be approximated by a power law at all. In the right 
column we show the extracted powers as a function of $\eta_f$, the scale 
parameter for the transverse flow. Since $R_\perp$ is not well represented
by a power law, the extracted slope depends somewhat on the fit region,
as indicated for the two sets of curves in Fig.~\ref{F11}b.
Altogether it is, however, clear that for pions the power $\alpha_\perp$ 
increases approximately linearly with $\eta_f$ and for kaons
somewhat more strongly. But even for large transverse flow rapidities 
$\eta_f \simeq 0.5$ the power remains below 0.2. In contrast, the 
corresponding power $\alpha_\parallel$ in a fit $R_\parallel(M_\perp) 
\propto M_\perp^{-\alpha_\parallel}$ is already 0.55 in the absence 
of transverse flow, reflecting the strong
boost-invariant longitudinal expansion. (Note that the decrease of 
$R_\parallel$ with increasing $M_\perp$ is faster than the 
$\sqrt{T/M_\perp}$-law suggested in Ref.~\cite{MS88} -- see also
Ref.~\cite{WSH96,SAT96} for a discussion of this point.) As the transverse 
flow is switched on, $\alpha_\parallel$ changes much more weakly than 
$\alpha_\perp$, showing that $R_\parallel$ is mostly sensitive to 
the longitudinal flow while $R_\perp$ is only affected by transverse
expansion. Again, kaons are affected by the transverse flow more
strongly than pions.

\section{Conclusions}
\label{sec5}

We have presented a numerical study of the Yano-Koonin-Podgoretski\u\i\ 
fit parameters for the two-particle correlation function. Our starting 
point were the recently derived model-independent expressions for the 
YKP parameters in terms of second order space-time variances of the source
emission function. These expressions allow for an easy evaluation of
the YKP parameters as functions of the pair momentum ${\bf K}$ and for
detailed parameter studies. We exploited them for a class of hydrodynamic
models describing locally thermalized and collectively expanding sources,
and we studied the dependence of the YKP parameters on the longitudinal and
temporal extension of the source as well as on its longitudinal
and transverse expansion velocity. 

In the context of such models it has been argued previously that in 
a certain approximation (which becomes exact in the absence of 
transverse $x$-$p$-correlations of the source
as, e.g., induced by transverse expansion flow) the YKP 
parametrisation achieves a perfect factorisation of the longitudinal 
and transverse spatial and the temporal extensions (``lengths of 
homogeneity'') of the source, in the comoving frame of the emitting fluid 
element. The velocity of this emitting fluid element is then given by 
the fourth YKP parameter (the YK velocity). Here we have shown numerically 
that, within these models, these features are preserved even in the
presence of transverse flow. The transverse radius parameter 
$R_\perp({\bf K})$ gives the effective transverse size of the source, 
the longitudinal radius parameter $R_\parallel({\bf K})$ its effective 
longitudinal size, both in the local rest frame of the emitter as seen 
by pairs with momentum ${\bf K}$. Also, the YKP parameter 
$R_0({\bf K})$ provides a direct estimate of the effective emission duration 
for particles with momentum ${\bf K}$; this estimate is quite accurate as 
long as the average transverse flow rapidity remains below 0.5 and the 
``lifetime parameter'' $\Delta\tau$ is not too small ($\Delta\tau > 1$ 
fm/$c$). 

We also showed analytically and numerically that the YK velocity obtained 
from the YKP fit 
is indeed approximately equal to the longitudinal velocity of the emitting
fluid element (the LSPS velocity), and that the small differences
between these two velocities can be understood quantitatively in terms
of asymmetries of the source around the point of maximum emissivity (its 
``saddle point''). This enables us to interpret the rise of the YK 
rapidity with the pair rapidity $Y$ as a direct consequence of the 
longitudinal expansion of our source, and the $M_\perp$-dependence of the 
slope of the function $Y_{_{\rm YK}}(Y)$ as a signature for the 
thermal smearing of the particle momenta in the fluid rest frame. 
In Ref.~\cite{HTWW96} we further showed that the slope of the function 
$Y_{_{\rm YK}}(Y)$ is nearly independent of the transverse flow rapidity 
$\eta_f$. Thus the YK rapidity is manifestly dominated by the 
{\em longitudinal} expansion and hardly affected by the transverse 
expansion at all. The latter causes, on the other hand, an 
$M_\perp$-dependence of the transverse radius parameter $R_\perp$, 
which in turn is completely unaffected by the longitudinal expansion. 
In the present model study, the YKP parametrisation thus 
not only leads to a factorisation of the (transverse and longitudinal) 
spatial and temporal aspects of the source {\em geometry}, but it also 
cleanly separates its transverse and longitudinal {\em dynamics}.

The sensitivity of the YKP radius parameters to the transverse expansion 
of the source was investigated quantitatively in Sec.~\ref{sec4d}.
While the longitudinal radius parameter $R_\parallel$ is affected very
little by the transverse flow (its strong $M_\perp$-dependence arises from
the strong longitudinal flow), the transverse radius shows a considerable
dependence on $\eta_f$ (but none to the longitudinal flow). Furthermore, 
transverse flow breaks the exact $M_\perp$-scaling of the YKP radius 
parameters which we showed to exist for $\eta_f=0$ (see also \cite{HTWW96a}). 
As explained in Sec.~\ref{sec4d}, both effects can be combined for 
a quantitative extraction of the mean transverse expansion velocity 
from the YKP radius parameters. The results from a comprehensive 
analysis \cite{WH96} of resonance decay contributions 
to the correlation function indicate that they don't jeopardise such 
a program.

We would not like to close without remarking that the class of source 
models studied here is restricted in one crucial aspect: if the 
source is ``opaque'' \cite{HV96}, (i.e.\ the particle emission is
strongly surface dominated rather than being distributed over the
whole emission region with a Gaussian geometric weight as assumed here),
the resulting differences between the variances $\langle \tilde x^2 \rangle$ 
and $\langle \tilde y^2 \rangle$ lead to much larger contributions in 
Eqs.~(\ref{1.17}), (\ref{1.19}) and (\ref{3.1}) than found in the present 
study. In Ref.~\cite{TH97} it is shown that this affects strongly the 
interpretation of measured YKP parameters, in particular of the
``temporal'' parameter $R_0^2$, which typically becomes strongly
negative for small $K_\perp$ due to the negative contribution from
$\langle \tilde x^2 - \tilde y^2\rangle$ in this case. However, that 
study also shows that presently available data on YKP radii from Pb+Pb 
collisions \cite{NA49} do not show any such indications for opaqueness 
of the source and favor models with volume dominated emission. The NA44 
data \cite{NA44QM96}
for HBT radii from heavy ion collisions with Pb targets at the CERN SPS,
which within the standard Cartesian parametrization seem to be consistent
with $R_o=R_s$, are more difficult to interpret because of the 
non-trivial shape of the acceptance window of this experiment in
the $Y-K_\perp$ plane and the lack of information on the $K$-dependence
of these parameters. Also, the important consistency relations (\ref{1.15}) 
so far have only been checked by the NA49 collaboration \cite{NA49} whose
data are in qualitative agreement with the numerical results presented
here \cite{H97}. A quantitative comparison with the experiments will be 
presented as soon as finalized data become available.

\acknowledgments
This work was supported by grants from DAAD, DFG, NSFC, BMBF and GSI. 
We gratefully acknowledge discussions with H. Appelsh\"auser, 
S. Chapman, D. Ferenc, P. Foka, M. Ga\'zdzicki, K. Kadija, H. Kalechofsky, 
M. Martin, P. Seyboth and C. Slotta. U.H. thanks the CERN Theory Group 
for their warm hospitality.

\appendix
\section{Space-time moments o the emission function}
\label{app}

Using cylindrical coordinates $\tau,\eta,r,\phi$ with $d^4x = 
\tau \, d\tau\, d\eta\, r\, dr\, d\phi$, we can write the emission function
(\ref{2.1}) as
 \begin{equation}
 \label{A1}
    S(x,K)\, d^4x = T(\tau) \, P(r,\phi) \, H(r,\eta)\, 
                    d\tau\, d\eta\, dr\, d\phi\, ,
 \end{equation}
with
 \begin{eqnarray}
 \label{A2}
    T(\tau) &=& {\tau \over \sqrt{2\pi (\Delta\tau)^2}}
       \exp \left(-{(\tau-\tau_0)^2\over 2(\Delta\tau)^2}\right)\, ,
 \\
 \label{A3}
    P(r,\phi) &=& {1\over 2\pi}\, e^{b(r)\cos\phi} \, ,
 \\
 \label{A4}
    H(r,\eta) &=& {r\over (2\pi)^2}\, 
                  \exp\left(-{r^2 \over 2 R^2}
                            -{(\eta-\eta_0)^2 \over 2(\Delta\eta)^2} 
                      \right)\,
                  M_\perp \cosh(\eta-Y)\, e^{-a(r)\cosh(\eta-Y)} \, ,
 \end{eqnarray}
where we defined
 \begin{eqnarray}
 \label{A5}
    a(r) &=& {M_\perp \over T} \cosh\eta_t(r)\, ,
 \\
 \label{A6}
    b(r) &=& {K_\perp \over T} \sinh\eta_t(r)\, .
 \end{eqnarray}
The $\phi$ and $\tau$ integrations can be done analytically. We use
 \begin{equation}
 \label{A7}
    \int_0^{2\pi} {d\phi \over 2 \pi}\, e^{b\cos\phi}\, \cos(n\phi) =
    I_n(b) 
 \end{equation}
and define
 \begin{eqnarray}
 \label{A8}
   T_0 &=& 
 \langle 1 \rangle_\tau = 
        \int_{-\infty}^\infty T(\tau)\, d\tau
        = \tau_0\, ,
 \\
 \label{A9}
   T_1 &=& 
 \langle \tau \rangle_\tau = 
        \int_{-\infty}^\infty \tau \, 
           T(\tau)\, d\tau  = \tau_0^2 + (\Delta\tau)^2 \, ,
 \\
 \label{A10}
   T_2 &=& 
 \langle \tau^2 \rangle_\tau = 
           \int_{-\infty}^\infty \tau^2\, 
           T(\tau)\, d\tau  = \tau_0^3 + 3\tau_0 (\Delta\tau)^2 \, .
 \end{eqnarray}
The variance of the $\tau$-distribution $T(\tau)$ is 
 \begin{equation}
 \label{A11}
    \langle \tau^2 \rangle_\tau - \langle \tau \rangle_\tau^2 =
    (\Delta\tau)^2 \left(1-\left({\Delta\tau \over \tau_0}\right)^2 \right)\, .
 \end{equation}
Defining further
 \begin{equation}
 \label{A12}
    \langle f(r,\eta)\rangle_* = 
    {\int_0^\infty dr \int_{-\infty}^\infty d\eta \, H(r,\eta)\, I_0(b(r))
     \, f(r,\eta) 
     \over
     \int_0^\infty dr \int_{-\infty}^\infty d\eta \, H(r,\eta)\, I_0(b(r))}
 \end{equation}
we find for the non-vanishing moments up to second order
 \begin{eqnarray}
 \label{A13}
   \langle x^2 \rangle &=& {1\over 2} \left\langle r^2 
                           \left(1 + {I_2(b(r))\over I_0(b(r))}\right)
                           \right\rangle_* \, ,
 \\
 \label{A14}
   \langle y^2 \rangle &=& {1\over 2} \left\langle r^2 
                           \left(1 - {I_2(b(r))\over I_0(b(r))}\right)
                           \right\rangle_* \, ,
 \\
 \label{A15}
   \langle z^2 \rangle &=& {T_2\over T_0} \left\langle \sinh^2\eta 
                                          \right\rangle_* \, ,
 \\
 \label{A16}
   \langle t^2 \rangle &=& {T_2\over T_0} \left\langle \cosh^2\eta 
                                          \right\rangle_* \, ,
 \\
 \label{A17}
   \langle x \rangle &=& \left\langle r\, {I_1(b(r))\over I_0(b(r))}
                         \right\rangle_* \, ,
 \\
 \label{A18}
   \langle z \rangle &=& {T_1\over T_0} \left\langle \sinh\eta 
                                        \right\rangle_* \, ,
 \\
 \label{A19}
   \langle t \rangle &=& {T_1\over T_0} \left\langle \cosh\eta 
                                        \right\rangle_* \, ,
 \\
 \label{A20}
   \langle xt \rangle &=& {T_1\over T_0} \left\langle r \cosh\eta \,
                          {I_1(b(r))\over I_0(b(r))} \right\rangle_* \, ,
 \\
 \label{A21}
   \langle zt \rangle &=& {T_2\over T_0} \left\langle \sinh\eta \cosh\eta 
                                         \right\rangle_* \, ,
 \\
 \label{A22}
   \langle xz \rangle &=& {T_1\over T_0} \left\langle r \sinh\eta \,
                          {I_1(b(r))\over I_0(b(r))} \right\rangle_* \, .
 \end{eqnarray}
For $\eta_f\ne 0$ the $\langle\dots\rangle_*$-averages have to be done 
numerically. For convenience we also give
 \begin{eqnarray}
 \label{A23}
   \langle \tilde x \tilde t \rangle &=& 
   {T_1\over T_0} \left[ 
   \left\langle r \cosh\eta\, {I_1(b(r))\over I_0(b(r))} \right\rangle_*
  -\left\langle r\, {I_1(b(r))\over I_0(b(r))} \right\rangle_*
   \left\langle \cosh\eta \right\rangle_*
   \right] \, ,
 \\
 \label{A24}
   \langle \tilde x^2 - \tilde y^2 \rangle &=& 
   \left\langle r^2\, {I_2(b(r))\over I_0(b(r))} \right\rangle_* 
  -\left\langle r\,   {I_1(b(r))\over I_0(b(r))} \right\rangle_*^2 \, .
 \end{eqnarray}
Please note that for small arguments $I_n(b) \sim b^n$. Thus for small 
$\eta_f$ and/or $K_\perp$, $\langle \tilde x \tilde t \rangle$
and $\langle \tilde x^2 - \tilde y^2 \rangle$ vanish linearly and 
quadratically, respectively. 


\end{document}